\documentclass[12pt]{article} 
\usepackage{epsfig,cite}
\usepackage{amsmath}
\usepackage{amssymb} 
\usepackage{color}
\usepackage{graphicx}
\usepackage{verbatim,mathrsfs,subfigure,feynmf}
\usepackage{url}
\usepackage{ytableau} 
\usepackage{cancel}
\usepackage{hyperref}
\usepackage{physics}
\usepackage[super]{nth}
\usepackage{slashed}
\usepackage{orcidlink}
\usepackage[toc,page]{appendix}
\usepackage{braket}
\usepackage{mathtools,slashed}
\usepackage[table]{xcolor} 
\usepackage{tikz}
\usepackage{tikz-feynman}
\tikzfeynmanset{warn luatex=false}
\tikzfeynmanset{compat=1.1.0}

\usepackage{cancel}
\usepackage{tcolorbox}

\usepackage{float}
\usepackage{xcolor}

\newcount\timecount
\newcount\hours \newcount\minutes  \newcount\temp \newcount\pmhours
\hours = \time
\divide\hours by 60
\temp = \hours
\multiply\temp by 60
\minutes = \time
\advance\minutes by -\temp
\def\hour{\the\hours}
\def\minute{\ifnum\minutes<10 0\the\minutes
            \else\the\minutes\fi}
\def\clock{
\ifnum\hours=0 12:\minute\ AM
\else\ifnum\hours<12 \hour:\minute\ AM
      \else\ifnum\hours=12 12:\minute\ PM
            \else\ifnum\hours>12
                 \pmhours=\hours
                 \advance\pmhours by -12
                 \the\pmhours:\minute\ PM
                 \fi
            \fi
      \fi
\fi
}

\def\monthname{\relax\ifcase\month 0/\or January\or February\or
   March\or April\or May\or June\or July\or August\or September\or
   October\or November\or December\else\number\month/\fi}

\def\bold#1{\setbox0=\hbox{$#1$}%
     \kern-.025em\copy0\kern-\wd0
     \kern.05em\copy0\kern-\wd0
     \kern-.025em\raise.0433em\box0 }

\def\beq{\begin{equation}}
\def\eeq{\end{equation}}

\def\ga{\mathrel{\raise.3ex\hbox{$>$\kern-.75em\lower1ex\hbox{$\sim$}}}}
\def\la{\mathrel{\raise.3ex\hbox{$<$\kern-.75em\lower1ex\hbox{$\sim$}}}}
\def\gev{{\rm \, Ge\kern-0.125em V}}
\def\tev{{\rm \, Te\kern-0.125em V}}
\def\gyr{{\rm \, G\kern-0.125em yr}}

\def\slash#1{\rlap{\hbox{$\mskip 1 mu /$}}#1}%

\def\gappeq{\mathrel{\rlap {\raise.5ex\hbox{$>$}}
{\lower.5ex\hbox{$\sim$}}}}
\def\lappeq{\mathrel{\rlap{\raise.5ex\hbox{$<$}}
{\lower.5ex\hbox{$\sim$}}}}
\def\Toprel#1\over#2{\mathrel{\mathop{#2}\limits^{#1}}}

\def\mpl{M_{\rm Pl}}

\def\m12{m_{1\!/2}}

 \newcommand{\mgrav}{m_{3/2}}
 \newcommand{\moh}{M_{1/2}}
 \newcommand{\gravrel}{\Omega_{3/2} h^2}

\def\mpl{M_{P}}

\def\bea{\begin{eqnarray}}
\def\eea{\end{eqnarray}}
\def\tb{\tan\beta}
 
 \newcommand{\GeV}{\; \mathrm{GeV}}
 \newcommand{\TeV}{\; \mathrm{TeV}}
\def\gm2{g_{\mu}-2}

 \newcommand{\Trp}{T_{\rm reh}^{\rm peak}}
 \newcommand{\Trpf}{T_{\rm reh}^{\rm peak,\phi}}

\def\m{\mu}

\makeatletter

\def\slash{\@ifnextchar[{\fmsl@sh}{\fmsl@sh[0mu]}}
\def\fmsl@sh[#1]#2{%
  \mathchoice
    {\@fmsl@sh\displaystyle{#1}{#2}}%
    {\@fmsl@sh\textstyle{#1}{#2}}%
    {\@fmsl@sh\scriptstyle{#1}{#2}}%
    {\@fmsl@sh\scriptscriptstyle{#1}{#2}}}
\def\@fmsl@sh#1#2#3{\m@th\ooalign{$\hfil#1\mkern#2/\hfil$\crcr$#1#3$}}
\makeatother

\def\beq{\begin{equation}}
\def\eeq{\end{equation}}

\newcommand{\suthree}{\ensuremath{\text{SU}(3)_{\text{c}}}}
\newcommand{\sutwo}{\ensuremath{\text{SU}(2)_{\text{L}}}}
\newcommand{\uone}{\ensuremath{\text{U}(1)_{\text{Y}}}}

\begin{document} 
\begin{titlepage}

\begin{center} 
\vspace*{1.5cm} 
 
 {\Large{\textbf{Gravitino Freeze-In Dark Matter with  \\ [2mm]
   an Additional Scalar Field}}} \\ 
 \vspace*{10mm} 
\end{center} 

\begin{center}
 {\bf  Georgios Georgilas,  Vassilis~C.~Spanos }   
\vspace{.7cm} 

{\it  National and Kapodistrian University of Athens, Department of Physics, \\
 Section of Nuclear {\rm \&} Particle Physics,  GR--15784 Athens, Greece} \\

\end{center}

\vspace{1.0cm}

\begin{abstract}
The gravitino is a prominent example of a freeze-in dark matter  candidate. Its relic abundance depends on the reheating temperature and on supersymmetry-breaking parameters, that is  the universal gaugino mass, $M_{1/2}$, and the gravitino mass, $m_{3/2}$. As a consequence, the reheating temperature consistent with the observed dark matter abundance exhibits a maximum value, $T_{\rm reh}^{\rm reak}$, which decreases as $M_{1/2}$ increases. This behavior gives rise to a tension between prospective lower bounds on the gluino mass from future collider searches and the high reheating temperatures required for successful thermal leptogenesis. In this work, we investigate a nonstandard cosmological scenario in which the thermal bath is supplemented by an additional scalar field. We show that, for a matter-like equation of state, this component can induce a substantial dilution of the gravitino abundance, thereby allowing significantly larger values of the reheating temperature. In contrast, for a kination-like equation of state, the gravitino abundance is enhanced rather than diluted, leading to a reduction of the maximum allowed reheating temperature.
\end{abstract}

\vspace{4cm}
\begin{flushleft}
\par\noindent\rule{8cm}{0.5pt}\\ 
{\small georgilasgeo@phys.uoa.gr, vspanos@phys.uoa.gr}
 \end{flushleft} 

\end{titlepage}

\baselineskip= 15.4  pt 

\newpage



\section{Introduction}
\label{sect:intro}

In the search for dark matter (DM), the so-called weakly interacting massive particle (WIMP) miracle has long been regarded as a particularly attractive possibility~\cite{Baltz:2006fm,Albertus:2026fbe}. One of the main reasons is that it offered the prospect of detecting the DM particle in collider experiments, such as those conducted at the Large Hadron Collider (LHC). This expectation arose from the observation that the DM pair-annihilation cross section required to reproduce the observed relic abundance via thermal freeze-out is typically of the order of weak-interaction cross sections, placing it within the potential reach of collider and direct-detection experiments.
Unfortunately, this expectation has not been realized, either in collider experiments or in direct DM searches, which aim to detect DM through its interactions with nucleons (or, at the parton level, with quarks) at approximately this interaction strength.

One way to evade these experimental and observational constraints is provided by the framework of freeze-in DM models~\cite{Hall:2009bx, Elahi:2014fsa, Bernal:2019mhf,Belanger:2018ccd}. In this scenario, the DM particle interacts so feebly with the Standard Model (SM)  sector that it never reaches thermal equilibrium in the early Universe. Instead, its relic abundance is gradually generated through rare scatterings and decays of particles in the thermal bath. Although the freeze-in mechanism has received considerable attention in recent years, the underlying idea is not new. Prominent early examples include the axion and the gravitino. 
In this paper we will focus on gravitino, but similar analysis can be applied  in a general freeze-in scenario. 

The gravitino can be produced in the early Universe after inflation primarily through two mechanisms: direct decays of the inflaton~\cite{Ellis:2015jpg, Garcia:2020wiy, Kallosh:1999jj, Giudice:1999am,Nilles:2001ry, Kawasaki:2006gs,Endo:2006qk,Dudas:2017rpa,Kaneta:2019zgw,Ellis:1982yb,Nanopoulos:1983up,Garcia:2017tuj,Kaneta:2023uwi} and thermal production from out-of-equilibrium processes~\cite{Eberl:2020fml,Eberl:2024pxr,Ellis:2015jpg,Rychkov:2007uq,Weinberg:1982zq,Ellis:1984eq,Khlopov:1984pf,Moroi:1993mb,Kawasaki:1994af,Moroi:1995fs,Ellis:1995mr,Bolz:1998ek,Bolz:2000fu,Steffen:2006hw,Pradler:2006qh,Pradler:2006hh,Pradler:2006tpx,Cheung:2011nn} involving particles in the thermal bath. The former mechanism is, in general, highly model dependent, as it depends sensitively on the details of the inflationary sector. By contrast, the latter provides a robust and generic mechanism that can account for the observed DM abundance. In this case, the resulting DM relic density depends predominantly on the reheating temperature.

The dependence of the gravitino DM abundance, $\gravrel$, on the reheating temperature has been extensively studied~\cite{Rychkov:2007uq,Eberl:2021gdf,Eberl:2024pxr}. In this relation, an additional supersymmetric parameter plays a crucial role: the universal gaugino mass, $M_{1/2}$. This dependence arises through the spin-$1/2$ goldstino components of the gravitino, whose interactions are enhanced for sufficiently small gravitino masses, $m_{3/2}$, relative to those of the transverse spin-$3/2$ modes.
Interestingly, for fixed values of $\Omega_{3/2} h^2$, $\moh$, and $\mgrav$, the reheating temperature exhibits a maximum. This peak occurs precisely in the transition region where the dominant contribution to gravitino production shifts from the spin-$1/2$ goldstino components to pure gravitino  spin-$3/2$ modes.

The position and value of $T_{\rm reh}^{\rm peak}$, under the assumption that gravitinos account for the total DM abundance, $\gravrel = 0.12$, have been investigated in details~\cite{Rychkov:2007uq,Eberl:2021gdf,Eberl:2024pxr}. 
It was found that $T_{\rm reh}^{\rm peak}$ decreases as the universal gaugino mass $M_{1/2}$ increases.
This behavior has important phenomenological and cosmological implications. In particular, for $M_{1/2} \simeq 1~\TeV$, which is approximately the current lower bound from the LHC~\cite{ATLAS:2017weo,
CMS:2019zmd}, one finds $T_{\rm reh}^{\rm peak} \simeq 10^9~\GeV$. A prospective future bound of $M_{1/2} \simeq 2~\TeV$ would reduce $T_{\rm reh}^{\rm peak}$ by approximately a factor of two.

The value of $T_{\rm reh}$ plays a central role in cosmology, particularly in the realization of successful baryogenesis via leptogenesis. In the case of standard thermal leptogenesis, the reheating temperature must typically satisfy $T_{\rm reh} \gtrsim 2 \times 10^9~\GeV$, in order to generate the observed baryon asymmetry of the Universe~\cite{Barbieri:1999ma,Abada:2006ea,Buchmuller:2004nz,Davidson:2008bu, Davidson:2002qv}.
Consequently, gravitino DM production through the freeze-in mechanism gives rise to a potential tension between collider constraints and cosmological requirements. Specifically, increasingly stringent lower bounds on the universal gaugino mass $M_{1/2}$ imply progressively smaller values of $T_{\rm reh}^{\rm peak}$, while successful thermal leptogenesis requires reheating temperatures at or above the scale quoted above. This tension becomes particularly significant if future experimental searches push the lower bound on $M_{1/2}$ to higher values.

It is important to note that any additional source of gravitino production would further reduce the allowed thermally produced contribution to the DM abundance and, consequently, would lower $T_{\rm reh}^{\rm peak}$. Therefore, a new framework is required in order to alleviate the aforementioned tension.
This can be achieved by introducing an additional component into the cosmological model, beyond radiation and gravitino DM. As we will show below, the presence of such a component can lead to a substantial dilution of the gravitino yield for fixed values of $m_{3/2}$ and $M_{1/2}$, thereby allowing for significantly larger values of $T_{\rm reh}^{\rm peak}$\footnote{An analogous effect, although in the context of the neutralino freeze-out DM, was studied in~\cite{Lahanas:2006xv,Lahanas:2011tk,Lahanas:2012hd}. In that case, the role of the additional field was played by  dilaton.}.

It is well known that the presence of the inflaton and its decays can lead to entropy dilution effects~\cite{Ellis:2015jpg,Giudice:1999am,Rychkov:2007uq,Gomez:2008js}, as can the presence of a saxion field~\cite{Co:2016fln,Co:2017orl}.
More generally, additional scalars  arise naturally in supergravity~\cite{Lahanas:1986uc,Ellis:2015kqa} and string-inspired frameworks, including moduli, flat directions, saxions, and hidden-sector scalars with suppressed couplings to the visible sector~\cite{Coughlan:1983ci,deCarlos:1993wie,Dine:1995kz,Co:2017orl}.
In the present work, however, we adopt a more general and model-independent approach and study the cosmological effect of an additional scalar component, $\phi$, not necessarily identified with either the inflaton or the saxion. The dynamics of this component are parametrized phenomenologically through three independent quantities: its initial energy density $\rho_\phi(T_{\rm reh})$, the equation-of-state parameter $w_\phi$, and its decay width $\Gamma_\phi$.

In our numerical analysis, we consider values of $w_\phi$ ranging from nearly zero, corresponding to a matter-like component, to unity, corresponding to a kination-dominated component. In the matter-like case, with $\Gamma_\phi = 10^{-12}~\GeV$, even a subdominant initial contribution, $r_\phi \equiv \rho_\phi/\rho_R = 10^{-3}$, can lead to a substantial dilution factor of order $10^2$. As a result, $T_{\rm reh}^{\rm peak}$ can increase by approximately two orders of magnitude. Smaller values of $\Gamma_\phi$ lead to an even larger dilution of the gravitino abundance.
In contrast, for the kination case ($w_\phi = 1$), we find dilution factors smaller than unity, implying an enhancement rather than a suppression of the gravitino abundance. Consequently, $T_{\rm reh}^{\rm peak}$ is reduced relative to its value in the standard cosmological scenario.

The paper is organized as follows. In Section~\ref{sect:std}, we briefly review the conventional freeze-in production of gravitino DM. In Section~\ref{sect:phi}, we introduce the nonstandard cosmological framework containing an extra  scalar  and derive the corresponding gravitino abundance. In Section~\ref{sect:num}, we present our numerical results, with particular emphasis on the dilution factor and the shifted reheating-temperature peak $T_{\rm reh}^{\rm peak,\phi}$. We also identify the regions of parameter space in which a substantial dilution of the gravitino abundance can be achieved. Finally, in Section~\ref{sect:recap}, we summarize our conclusions and discuss possible directions for future work.

\section{Gravitino  DM and reheating temperature}
\label{sect:std}

In this  section we will review the conventional gravitino thermal production details. The key ingredient for calculating 
the gravitino abundance $Y_{3/2}$ is the production rate $\gamma_{3/2}$.    For this we will 
use the results from~\cite{Eberl:2020fml}. In the following, for the sake of simplicity, we retain only the dominant contribution associated with the soft supersymmetry (SUSY) breaking gaugino masses, neglecting the contribution from the trilinear top-quark coupling parameter, $A_t$, since its effect is numerically subdominant~\cite{Rychkov:2007uq,Eberl:2020fml, Eberl:2024pxr}.

Following the parametrization of the $\gamma_{3/2}$ as in~\cite{Eberl:2024pxr} we have
\beq 
\gamma_{3/2} = \frac{T^6}{\mpl ^2}  \frac{3  \zeta(3)\,}{16  \pi^3 }   \sum_{N = 1}^3  c_N \, g_N^2 \,  \left(1 + \frac{M^2_N}{3 m^2_{3/2}}\right) \ln\left( \frac{k_N}{g_N}\right)  \,,
\label{eq:gammatotpar}
\eeq
where the factors  $c_N$ ,  $k_N$ are taken from Table \ref{table:param} and the sum runs over the three Standard Model (SM) gauge groups.

\begin{table}[h!]
\centering
\begin{tabular}{|c || c |c| }
\hline 
SM group &  $c_N$ & $k_N$   \\
\hline  \hline  \\ [-6.5mm]
\uone & $ 35.56 $ & $0.85$ \\
\sutwo & $ 35.33 $  & $1.38$\\ 
\suthree & $ 29.40 $  & $3.07$ \\
\hline
\end{tabular}
\caption{The values of the constants $c_N, k_N$   that parametrize  the $\gamma_{3/2}$ in Eq.~(\ref{eq:gammatotpar}),  as given in~\cite{Eberl:2024pxr}. 
Each value corresponds to a particular SM gauge group, \uone , \sutwo, and \suthree . }
\label{table:param}
\end{table}

Here, we assume a unified gauge coupling,
$g_{\rm GUT} = \sqrt{\pi/6}$,
at the Grand Unified (GUT) scale. We also parametrize the gaugino masses as
$M_N = M_{1/2}\,{g_N^2}/{g_{\rm GUT}^2}$,
where the gauge couplings $g_N$ are evaluated at the relevant energy scale, identified here with the temperature. The running of the gauge couplings with the energy scale is assumed to follow that of the Minimal Supersymmetric Standard Model (MSSM).

In order to obtain the thermal gravitino abundance one has to solve 
the Boltzmann equation for the gravitino number density $n_{3/2}$ 
\beq
\frac{{\rm d} {n}_{3/2}}{{\rm d}t} + 3H n_{3/2} = \gamma_{3/2}\,,
\label{eq:boltzmann}
\eeq
where $H$ is the Hubble constant. The gravitino abundance $Y_{3/2}$ is defined as
\beq
Y_{3/2}= \frac{n_{3/2}}{s}\,,
\label{eq:abundance}
\eeq
where we have used for the entropy density  $s=(2\pi^2/45)g_{*\rm s}T^3$. Substituting $Y_{3/2}$
into~(\ref{eq:boltzmann}) we get  the Boltzmann equation for $Y_{3/2}$
\beq
\frac{{\rm d} Y_{3/2}}{{\rm d}t}  =  \frac{\gamma_{3/2}}{ s}\, .
\label{eq:Y_boltz}
\eeq
Using the  entropy conservation relation
\beq
\frac{{\rm d}( g_{*s} T^3a^3)}{{\rm d}t}=0 \, , 
\eeq 
where  $g_{*s}$ are the effective entropy degrees of freedom and $a$ the scale factor,  we can trade time with temperature in Eq.~(\ref{eq:Y_boltz})
as
\beq
\frac{{\rm d}Y_{3/2}}{{\rm d}T}  = 
       -\frac{  \gamma_{3/2}  } {H\, T\,  s} \left[1+\frac{T}{3} \frac{{\rm d}\ln g_{*s}  }{{\rm d}T}\right]\,.
\label{eq:Y_boltz2}
\eeq
By integrating  Eq.~(\ref{eq:Y_boltz2}) 
from  $T_{\rm reh}$ down  to temperature $T$ and assuming 
  that inflaton decay and thermalization are instantaneous and simultaneous at $T_{\rm reh}$, 
one gets   
\bea
Y_{3/2}(T)&=&Y_{3/2}(T_{\rm reh})  - \int_{T_{\rm reh}}^T {\rm d}\tau \, \frac{\gamma_{3/2} (\tau)} { H(\tau)\, s(\tau)\,\tau}\left[1+\frac{\tau}{3} \frac{{\rm d}\ln g_{*s}(\tau)}{{\rm d}\tau}\right]\,.
\label{eq:Y_integr}
\eea
Furthermore, applying the standard freeze-in scenario assuming negligible initial  gravitino 
abundance at $T_{\rm reh}$,    $Y_{3/2}(T_{\rm reh})\simeq 0$,   and ignoring 
a weak temperature dependence of the integrand in the r.h.s. of Eq.~(\ref{eq:Y_integr}), 
 we get that the gravitino abundance in the limit $T \ll T_{\rm reh}$ reads as
\beq
Y_{ 3/2}(T)   \underset{T \ll T_{\rm reh}}{\longrightarrow}    \frac{\gamma _{3/2}(T_{\rm reh})}{H(T_{\rm reh}) \,\,  s(T_{\rm reh})}\, .
\label{eq:abundance_app}
\eeq
\begin{figure}[t!]
\centering
\includegraphics[width=0.48\textwidth]{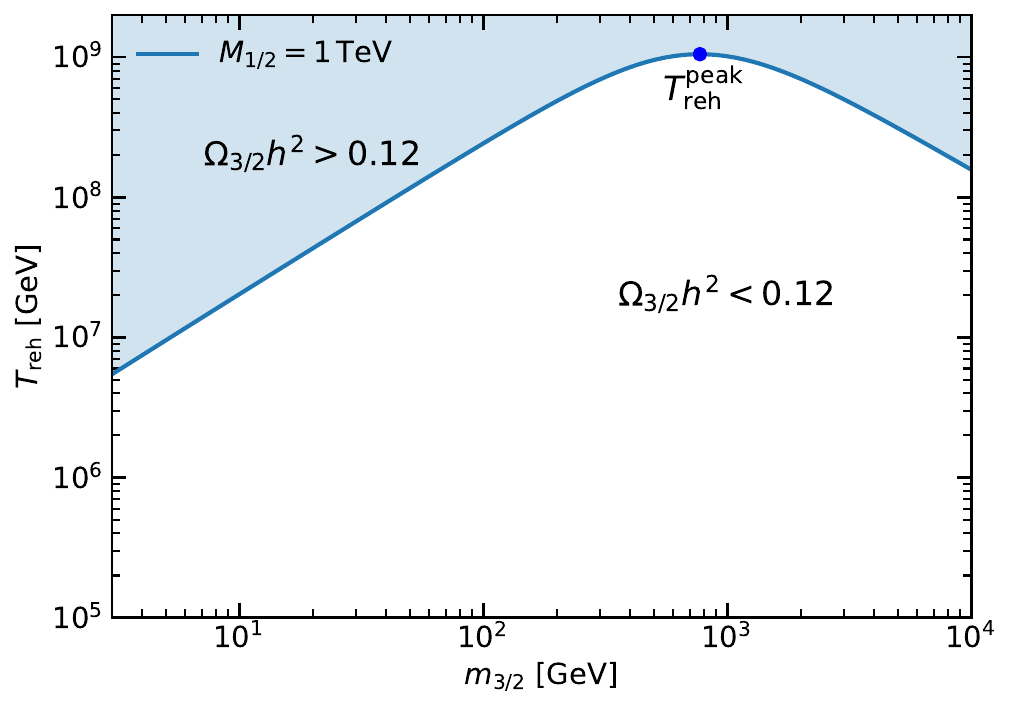}
\includegraphics[width=0.48\textwidth]{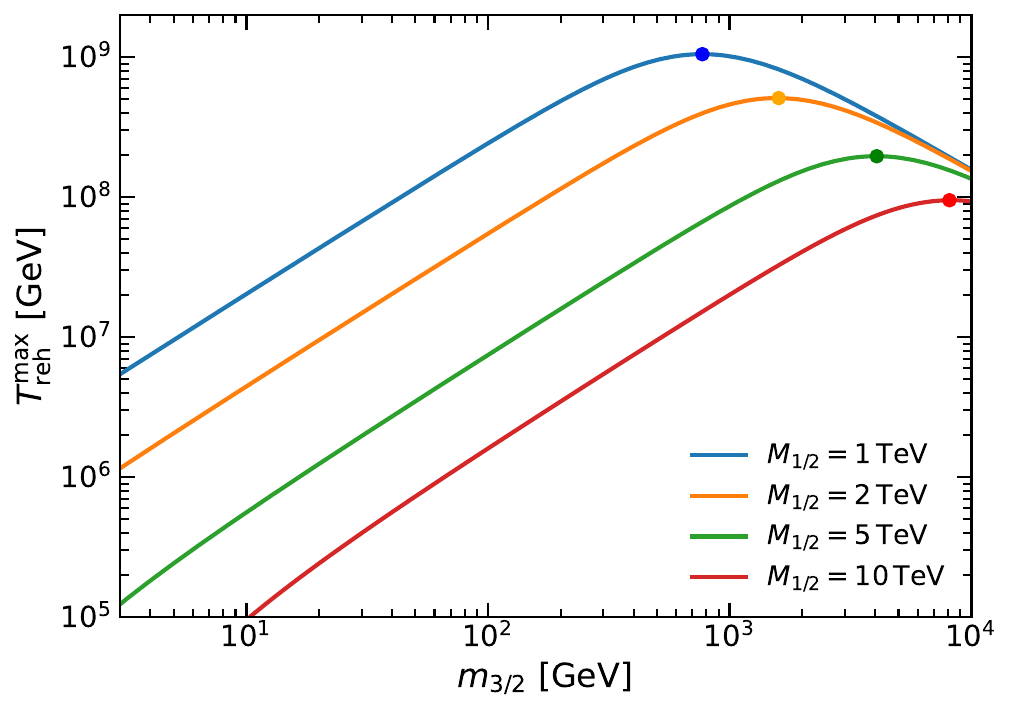}
\caption{\it  In the left panel, we plot the contour corresponding to $\gravrel = 0.12$ for $M_{1/2} = 1~\TeV$. The shaded region is cosmologically excluded, as it corresponds to $\gravrel > 0.12$. The unshaded region below the contour allows for additional contributions to the DM abundance. 
In the right panel, we show the corresponding contours for $\moh = 1$, $2$, $5$, and $10~\TeV$.
The points at the top of the curves indicate $\Trp$, namely the maximum reheating temperature consistent with the observed DM abundance for a given value of $\moh$. The corresponding values are listed in Table~\ref{table:Tehhmax}.}
\label{fig:grav_conv}
\end{figure}
Moreover, using the definitions 
\beq
\Omega _{3/2} =\frac{\rho_{3/2}}{ \rho _{\rm cr}}\,   , \qquad  \rho_{3/2} = n_{3/2}  m_{3/2}\,,
\label{eq:definitions}
\eeq
 and the Eqs~(\ref{eq:abundance}) and (\ref{eq:abundance_app}),  one  gets $\Omega _{3/2} h^2 $ as
 \beq
\Omega _{3/2} h^2 = \frac{\rho_{3/2}(t_0) h^2}{\rho _{\rm cr}} = \frac{m_{3/2} Y_{3/2}(T_0)s(T_0) h^2}{\rho _{\rm cr}}
  \simeq  1.33\times 10^{24} \frac{m_{3/2} \gamma_{3/2}(T_{\rm reh})}{T_{\rm reh}^5} \,.
  \label{eq:dmdensity}
\eeq
Finally, using Eq.~(\ref{eq:gammatotpar}), one obtains the following convenient approximate expression for evaluating $\gravrel$
 \beq  
 \Omega _{3/2} h^2 \simeq  0.016 \left( \frac{m_{3/2}}{1\TeV}\right) \frac{T_\mathrm{reh}}{10^{10} \GeV}  \sum_{N = 1}^3 \, c_N \, g_N^2 \,  \left(1 + \frac{M^2_N}{3 m^2_{3/2}}\right) \ln\left( \frac{k_N}{g_N}\right) \,. 
\label{eq:dmdensity_app}
\eeq

We have verified numerically that the analytic approximation given in Eq.~(\ref{eq:dmdensity_app}) agrees with the exact result obtained from Eq.~(\ref{eq:Y_integr}) at the level of better than $1\%$.
We recall that, in the above calculations, $\rho_{\rm cr} = 3 H_0^2 M_{\rm Pl}^2$ denotes the critical energy density, while $H_0 = 100\, h\, {\rm km\, s^{-1}\, Mpc^{-1}}$ is the present Hubble parameter. We also use $T_0 = 2.7~{\rm K}$ for the current temperature of the cosmic microwave background. The entropy degrees of freedom at the relevant temperatures are given by
$g_{*s}(T_0)=\frac{43}{11}$ and $g_{*s}(T_{\rm reh})=\frac{915}{4}$.
The latter value corresponds to the effective number of relativistic degrees of freedom in the   MSSM relevant for the determination of $H(T_{\rm reh})$.

According to the latest data from the Planck Collaboration, the observed DM abundance in the Universe is
$\gravrel = 0.12 \pm 0.0012$~\cite{Planck:2018vyg}. Assuming that the thermally produced gravitino abundance accounts for the entirety of the observed DM, one can derive constraints on the gravitino mass and/or the reheating temperature.
As we will demonstrate, however, these constraints can be substantially relaxed in the presence of an extra scalar field coexisting with gravitinos during the freeze-in production era.

In Fig.~\ref{fig:grav_conv} we plot curves with fixed $\gravrel=0.12$ in the $m_{3/2},T_{\rm reh}$ plane. For plotting these curves we have used  the exact numerical 
solution of the gravitino abundance Boltzmann Eq.~(\ref{eq:Y_boltz2}). As explained earlier, identical results were obtained using the approximation in Eq.~(\ref{eq:dmdensity_app}).
As can be seen in this figure, for fixed values of $M_{1/2}$ the reheating temperature exhibits a maximum value, denoted by $T_{\rm reh}^{\rm peak}$, occurring at a particular value of the gravitino mass $m_{3/2}$. The origin of this behavior was discussed in~\cite{Eberl:2024pxr} and can be understood directly from Eq.~(\ref{eq:dmdensity_app}).

For fixed $\Omega_{3/2} h^2$ and $M_{1/2}$, Eq.~(\ref{eq:dmdensity_app}) determines the dependence of $T_{\rm reh}$ on the gravitino mass. In the regime $M_{1/2} \lesssim 3 \TeV$, $m_{3/2} \lesssim 1 \TeV$,
the contribution proportional to ${M_N^2}/{3 m_{3/2}^2}$ dominates the gravitino production rate. In this region, increasing $m_{3/2}$ suppresses the goldstino-enhanced contribution, and therefore larger values of $T_{\rm reh}$ are required in order to reproduce the observed DM abundance. As a result, $T_{\rm reh}$ increases with increasing $m_{3/2}$.

\begin{table}[t!]
\begin{center}
$\begin{array}{ |c||c|c|    }
\hline
M_{1/2}\, (\TeV) &  m_{3/2} \, (\GeV)  &  T_{\rm reh}^{\rm peak} \, (\GeV)       \\
\hline
\hline
1  & 771    &      1.05 \times  10^{9}       \\
\hline
2  &     1596     &    5.10 \times  10^{8}       \\
\hline
5  &    4065    &     1.96 \times  10^{8}       \\
\hline
10  &   8124     &      9.51 \times  10^{7}       \\
\hline
\end{array}$
\caption{\it The  $m_{3/2}$ and $T_{\rm reh}^{\rm peak}$  values for various  $M_{1/2}$ cases,  assuming complete gravitino DM, that is  $\Omega_{3/2}h^2=0.12$. These peak values correspond in cases in Fig.~\ref{fig:grav_conv}. We will be using   $M_{1/2}=1,2,5,10 \TeV$, as benchmark values  in our phenomenological analysis. }
\label{table:Tehhmax}
\end{center}
\end{table}

For sufficiently large gravitino masses, however, the term $M_N^2/(3m_{3/2}^2)$ becomes subdominant compared to unity. In this regime, the transverse spin-$3/2$ gravitino modes dominate the production process, and Eq.~(\ref{eq:dmdensity_app}) implies an approximately inverse dependence of $T_{\rm reh}$ on $m_{3/2}$. Consequently, $T_{\rm reh}$ starts decreasing as $m_{3/2}$ increases further. The transition between these two regimes gives rise to the peak structure observed in Fig.~\ref{fig:grav_conv}.
For larger values of $M_{1/2}$, the ratio $M_N^2/(3m_{3/2}^2)$ remains dominant over a broader range of gravitino masses. In this case, the peak gradually disappears and the reheating temperature becomes approximately proportional to $m_{3/2}$. 

The features discussed above are displayed in Fig.~\ref{fig:grav_conv}. In the left panel, we show the contour defined by $\gravrel = 0.12$ for the representative case $\moh = 1\TeV$. The region above the curve corresponds to an overproduction of gravitino DM and is therefore cosmologically excluded. By contrast, the region below the contour remains phenomenologically viable, since it allows for either partial gravitino DM or additional non-gravitino contributions to the total DM abundance. The highest point along the contour identifies $\Trp$, namely the largest reheating temperature compatible with the observed gravitino relic abundance for the chosen value of $\moh$.

The right panel illustrates the evolution of the $\gravrel = 0.12$ contour as the universal gaugino mass increases. In particular, we present the cases $\moh = 1$, $2$, $5$, and $10~\TeV$, emphasizing the gradual shift of $\Trp$ toward lower values as $\moh$ increases. As discussed previously, the benchmark value $\moh = 1~\TeV$ lies only a few hundred GeV above the current lower bounds from ATLAS Collaboration and CMS Collaboration searches~\cite{ATLAS:2017weo,CMS:2019zmd}.
On the other hand, within the framework of the Constrained Minimal Supersymmetric Standard Model (CMSSM), values as large as $\moh = 10~\TeV$ may be favored, particularly for moderate values of $\tb \sim 10$, once various direct and indirect phenomenological constraints are taken into account~\cite{Ellis:2022emx}.

For completeness, Table~\ref{table:Tehhmax} summarizes the corresponding values of $m_{3/2}$ and $\Trp$ for several representative choices of $\moh$, under the assumption that gravitinos account for the entire DM abundance, $\Omega_{3/2} h^2 = 0.12$. These parameter points coincide with the maxima appearing in Fig.~\ref{fig:grav_conv}. In the phenomenological analysis that follows, we will employ as benchmark choices the values $\moh = 1$, $2$, $5$, and $10\TeV$. All these points are indicated at the maxima of the curves in Fig.~\ref{fig:grav_conv}. In the following analysis, they will be used as benchmark points to investigate the effects induced by the presence of the additional $\phi$ component, namely the dilution or enhancement of the gravitino abundance.

\section{Gravitino freeze-in in the presence of  $\boldsymbol{\phi}$ }
\label{sect:phi}
We extend the standard radiation-dominated treatment of thermal gravitino production by introducing an additional energy component, $\rho_\phi$. Within supergravity and string-inspired frameworks, the appearance of extra scalar degrees of freedom during the thermal history of the Universe is quite natural. 
More generally, $\phi$ may represent any long-lived scalar condensate whose energy density becomes cosmologically relevant after reheating and before the onset of the standard radiation-dominated era. Depending on its equation of state and decay properties, this component can modify the expansion history of the Universe, generate entropy production, and consequently affect the freeze-in production of gravitinos.

This work does not address the microscopic origins or dynamical nature of $\phi $.
We therefore model it phenomenologically as a decaying fluid characterized by an equation-of-state parameter $w_\phi$, a decay width $\Gamma_\phi$, and an initial energy density $\rho_\phi(T_{\rm reh})$. Accordingly, the cosmological framework considered here consists of three components: radiation, gravitinos, and the scalar component $\phi$.

Since no specific scalar potential $V(\phi)$ is assumed, we do not impose the relation
\beq
w_\phi = \frac{P_\phi}{\rho_\phi} =
\frac{ \dot{\phi}^2/2 - V(\phi)}
{ \dot{\phi}^2/2+ V(\phi)} \, .
\eeq
Instead, $w_\phi$ is treated as a free phenomenological parameter. In particular, we consider the range
$0 \leq w_\phi \leq 1$, interpolating between matter-like behavior ($w_\phi = 0$) and a kination-dominated phase ($w_\phi = 1$), while also paying special attention to the radiation-like case, $w_\phi = 1/3$.

The cosmological evolution of the system is governed by
\begin{align}
\dot{\rho}_\phi &=-3(1+w_\phi)H\, \rho_\phi-\Gamma_\phi\, \rho_\phi \,, \nonumber \\
\dot{\rho}_R &=-4H\, \rho_R+\Gamma_\phi\, \rho_\phi \,, \nonumber \\
\dot{\rho}_{3/2} &=-3H\, \rho_{3/2}+m_{3/2}\,\gamma_{3/2}(T)\,,
\label{eq:eomphi}
\end{align}
where the Hubble parameter is determined through
\begin{equation}
H^2=\frac{1}{3M_P^2}\left(\rho_\phi+\rho_R\right)\,.
\label{eq:Hphi}
\end{equation}
Since throughout the evolution one has
$\rho_{3/2} \ll \rho_\phi,\rho_R$,
the gravitino contribution to the total energy density can safely be neglected in the Friedmann equation. 
The thermal gravitino production rate  $\gamma_{3/2}(T)$, is given by Eq.~(\ref{eq:gammatotpar}).

The radiation energy density and entropy density are related to the temperature through
\begin{equation}
\rho_R(T)=
\frac{\pi^2}{30}\,g_{*}(T)\,T^4\,
\end{equation}
and
\begin{equation}
s(T)=
\frac{2\pi^2}{45}\,g_{*s}(T)\,T^3\,, 
\end{equation}
respectively.

Because the decay of $\phi$ continuously injects entropy into the thermal bath, the use of cosmic time or temperature as the independent evolution variable is not particularly convenient. Instead, it is advantageous to parametrize the evolution in terms of the number of e-folds, $N = \ln a$, 
where $a$ denotes the scale factor. Therefore in terms of $N$, Eqs.~(\ref{eq:eomphi}) become
\begin{align}
\frac{d\rho_\phi}{dN}&=-3(1+w_\phi)\rho_\phi
-\frac{\Gamma_\phi}{H}\rho_\phi ,\nonumber \\
\frac{d\rho_R}{dN}&=-4\rho_R+\frac{\Gamma_\phi}{H}\rho_\phi ,\nonumber \\
\frac{d\rho_{3/2}}{dN}&=-3\rho_{3/2}+\frac{m_{3/2}\gamma_{3/2}(T)}{H}\, .
\end{align}

The numerical evolution starts at $N=0$, corresponding to the reheating temperature $T_{\rm reh}$. The initial conditions are chosen as
\begin{align}
&\rho_R(0)=\frac{\pi^2}{30}g_{*}(T_{\rm reh})T_{\rm reh}^4\,, \nonumber \\
&\rho_\phi(0)=r_\phi\,\rho_R(0)\,,\nonumber \\
&\rho_{3/2}(0)=0\,,
\end{align}
where the parameter
\begin{equation}
r_\phi \equiv\left.\frac{\rho_\phi}{\rho_R}\right|_{T=T{\rm reh}}
\label{eq:rho}
\end{equation}
measures the initial abundance of the scalar component relative to radiation at reheating. The system is then evolved numerically from $T_{\rm reh}$ down to the present temperature $T_0$.

\begin{figure}[t!]
\centering
\includegraphics[width=0.48\textwidth]{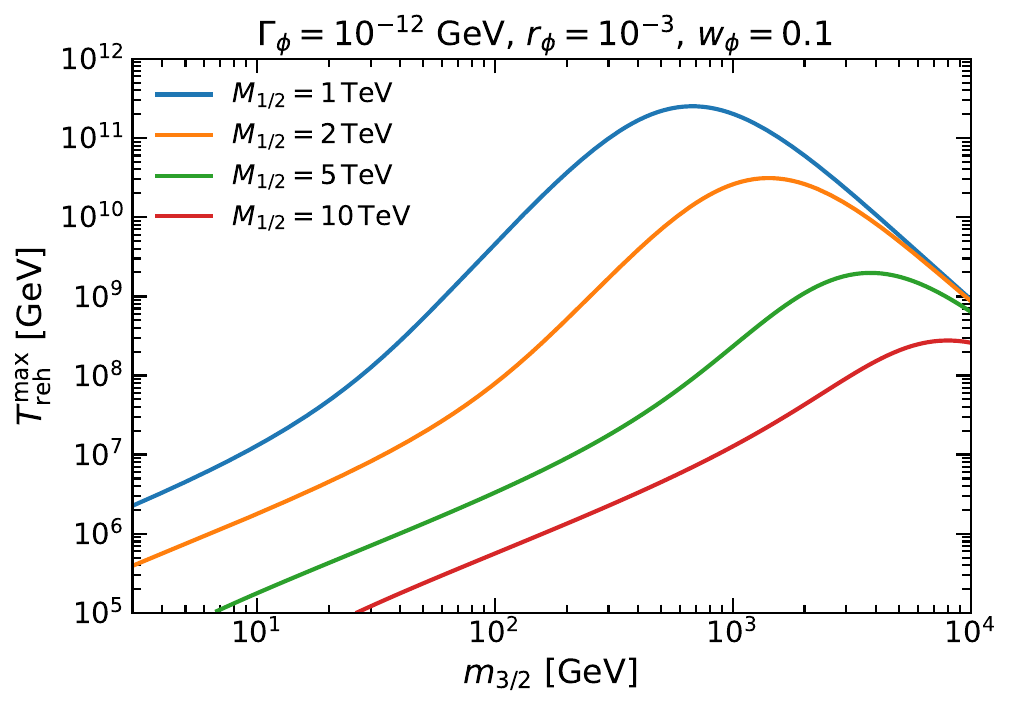}
\caption{\it  The curves of  fixed $\Omega_{3/2}h^2=0.12$ in the $m_{3/2},T_{\rm reh}^{\rm max}$ 
plane   in the presence of $\phi$ component,
for  $r_\phi=10^{-3}$, $w_\phi=0.1$ and $\Gamma_\phi=10^{-12} \GeV$.}
\label{fig:grav_ns}
\end{figure}

The final value of the evolution variable, $N_{\rm final}$, is not fixed beforehand but is instead determined dynamically during the numerical integration.
At each integration step, the temperature is reconstructed from the radiation energy density according to
\begin{equation}
\rho_R(N)=\frac{\pi^2}{30}\,g_{*}(T)\,T^4.
\end{equation}
The integration is terminated once the condition $T(N_{\rm final}) = T_{\rm final}$
is satisfied, where $T_{\rm final}$ is chosen sufficiently small, typically
$T_{\rm final}\sim 10^{-4}~\GeV$,
so that gravitino production has effectively ceased and the entropy injection from $\phi$ decays has been completed.

We note that in the standard radiation-dominated case, assuming constant relativistic degrees of freedom, one has
$T \sim a^{-1}$, and therefore 
\begin{equation}
N \simeq
\ln\left(
\frac{T_{\rm reh}}{T}
\right),
\end{equation}
and consequently  
\begin{equation}
N_{\rm final}^{\rm (RD)}
\simeq
\ln\left(
\frac{T_{\rm reh}}{T_{\rm final}}
\right).
\end{equation}
In the presence of the additional $\phi$ component, however, this relation is modified due both to the altered expansion rate and to the entropy injection associated with $\phi$ decays. Consequently, the relation between temperature and scale factor departs from the standard radiation-dominated behavior, and the value of $N_{\rm final}$ must be determined numerically, as done throughout our analysis.

After the numerical integration, the final gravitino yield is computed as
\begin{equation}
Y_{3/2}^{\phi}
=
\frac{n_{3/2}}{s}
=
\frac{\rho_{3/2}}{m_{3/2}\, s(T)}\, .
\end{equation}
The relic density is then
\begin{equation}
\Omega_{3/2}^{\phi} h^2
=
2.742\times 10^8
\left(\frac{m_{3/2}}{\rm GeV}\right)
Y_{3/2}^{\phi}\, .
\end{equation}
\begin{table}[t!]
\begin{center}
$\begin{array}{ |c||c|c| r |    }
\hline
M_{1/2}\, (\TeV) &  m_{3/2} \, (\GeV)  &  \Trpf \, (\GeV) & \Delta_\phi     \\
\hline
\hline
1  & 675     &     2.51 \times  10^{11}  & 187.6       \\
\hline
2  &     1472     &     3.11 \times  10^{10} &    49.6     \\
\hline
5  &      3856    &      1.97 \times  10^{9}  &  8.7  \\
\hline
10  &    7777    &     2.76 \times  10^{8}  &    2.6     \\
\hline
\end{array}$
\caption{\it The  $m_{3/2}$ and $T_{\rm reh}$ peak values for various  $M_{1/2}$ cases,  assuming   $\Omega_{3/2}h^2=0.12$, in in the presence of $\phi$ component, using $r_\phi=10^{-3}$, $w_\phi=0.1$ and $\Gamma_\phi=10^{-12} \GeV$.
 These peak values correspond in cases in Fig.~\ref{fig:grav_ns}. 
 We give also the dilution factor $\Delta_\phi$ for these cases defined
 as $\Omega_{3/2} /\Omega_{3/2}^{\phi}$ at the position of the new peak.}
\label{table:Tehhmax_phi1}
\end{center}
\end{table}
For comparison, we also compute the standard radiation-dominated result
$\Omega_{3/2} h^2$, that  correspond to the result
presented in Eq.~(\ref{eq:dmdensity})\footnote{In the following, we reserve the symbols $Y_{3/2}$ and $\Omega_{3/2} h^2$ for the gravitino yield and the corresponding DM relic abundance calculated in the standard radiation-dominated scenario discussed in Section~\ref{sect:std}. By contrast, the quantities $Y_{3/2}^{\phi}$ and $\Omega_{3/2}^{\phi} h^2$ will denote the corresponding quantities calculated in the presence of the additional $\phi$ component.}. 
This way  one can  define the dilution factor due 
to the presence of $\phi$
\begin{equation}
\Delta_\phi
=\frac{\Omega_{3/2} }{\Omega_{3/2}^{\phi} }\, .
\label{Eq:delta}
\end{equation}
A value $\Delta_\phi>1$ corresponds to a suppression of the gravitino abundance relative to the standard cosmological scenario, whereas 
$\Delta_\phi<1$ corresponds to an enhancement.

As a first step toward presenting our results in the presence of the additional $\phi$ component, we construct the analogue of Fig.~\ref{fig:grav_conv}. In Fig.~\ref{fig:grav_ns}, we plot the contours corresponding to fixed $\Omega_{3/2} h^2 = 0.12$ in the $(m_{3/2},T_{\rm reh})$ plane, obtained by solving the gravitino Boltzmann equation in the nonstandard cosmological background described above.
For illustration, we choose the representative values
$ r_\phi = 10^{-3}$, $w_\phi = 0.1$, $\Gamma_\phi = 10^{-12}~\GeV,$
and consider the benchmark cases
$\moh = 1, 2,5$ and 10 TeV.
We observe that, for $\moh = 1~\TeV$, the peak reheating temperature is enhanced by nearly two orders of magnitude relative to the standard radiation-dominated scenario. The enhancement becomes progressively smaller for larger values of $\moh$.

In Table~\ref{table:Tehhmax_phi1}, we list the corresponding peak values, denoted by $\Trpf$, for the benchmark choices of $\moh$. Comparison with Table~\ref{table:Tehhmax} clearly illustrates the enhancement of the peak reheating temperatures induced by the presence of the $\phi$ component. In the last column, we also provide the corresponding dilution factor, $\Delta_\phi$, calculated from Eq.~(\ref{Eq:delta}).

\begin{figure}[t!]
\centering
\includegraphics[width=0.48\textwidth]{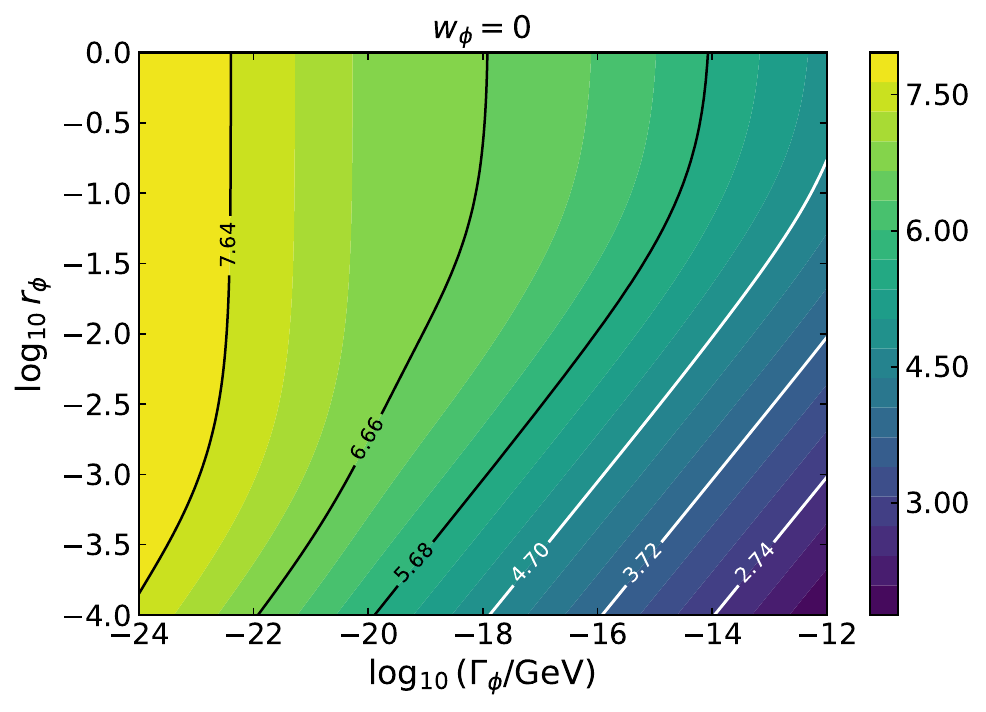}
\includegraphics[width=0.48\textwidth]{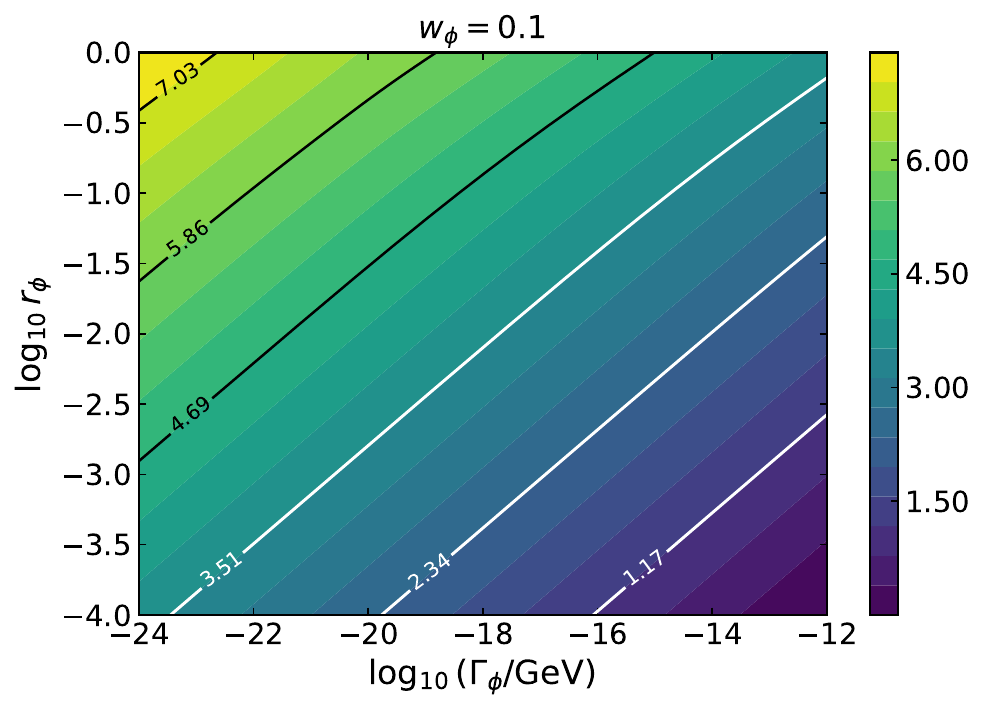}\\[0.5cm]
\includegraphics[width=0.49\textwidth]{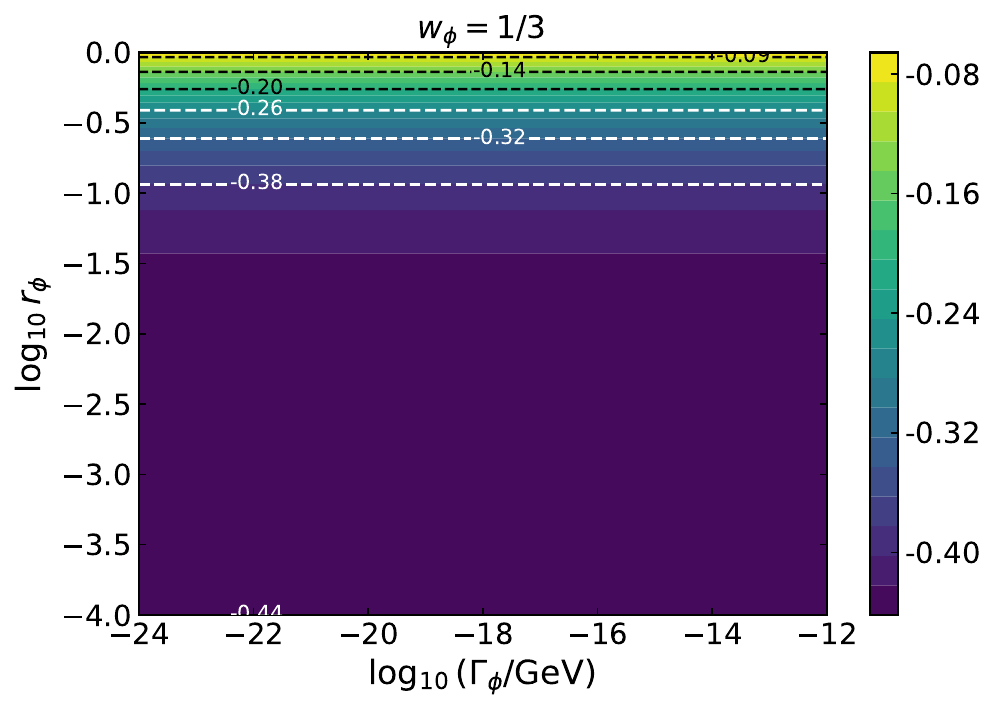}
\includegraphics[width=0.49\textwidth]{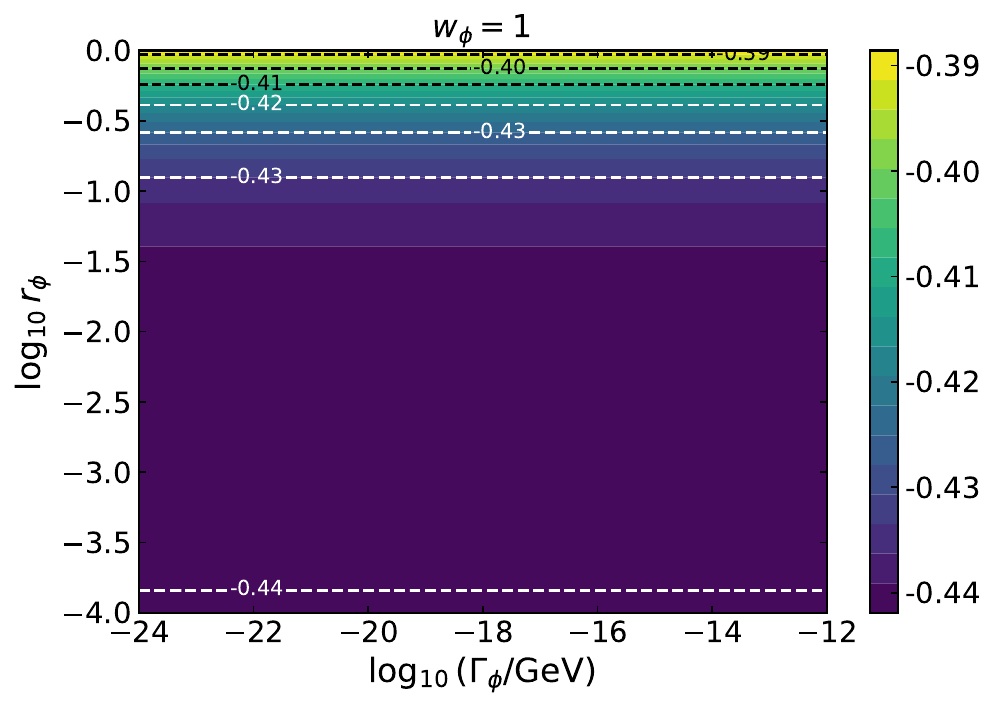} 
\caption{\it  The dilution factor $\Delta_\phi$ is shown for $w_\phi = 0$, $0.1$, $1/3$, and 1, in the $(\Gamma_\phi, r_\phi)$ plane. In each panel, we present contours of $\log \Delta_\phi$. In all cases, we fix the universal gaugino mass to $M_{1/2}=1~\TeV$ and the dilution factor is calculated at the $\Trp$,  for $\mgrav=771 \GeV$ as in Table~\ref{table:Tehhmax}. Solid contours correspond to positive values, while dashed contours correspond to negative ones.}
\label{fig:dil_1tev}
\end{figure}

In Fig.~\ref{fig:dil_1tev}, we present the dilution factor $\Delta_\phi$ in the $(\Gamma_\phi,r_\phi)$ plane for the representative values $w_\phi = 0$, $0.1$, $1/3$, and $1$. In all cases, we fix the universal gaugino mass to $M_{1/2}=1\TeV$. The dilution factor is evaluated at the standard reheating peak, $\Trp$, corresponding to  $\mgrav=771 \GeV$ as in Table~\ref{table:Tehhmax}.
The reason for evaluating $\Delta_\phi$ at the standard peak is to illustrate more clearly the magnitude of the dilution effect, particularly in the matter-like regime. As will be shown in the following section, for $w_\phi=0$ the resulting shifted reheating peak, $\Trpf$, is pushed to values exceeding $10^{16}\GeV$, outside the phenomenologically allowed range adopted in our analysis.

The choice $w_\phi = 0.1$ is included in order to study a case close to matter domination while avoiding the extremely large dilution effects obtained for the exact matter-like case, $w_\phi = 0$. 
In each panel, we display contours of $\log \Delta_\phi$, for  $M_{1/2}=1~\TeV$. The decay width is varied in the range $10^{-24}\GeV \lesssim \Gamma_\phi \lesssim 10^{-12}\GeV$, with the lower end corresponding to lifetimes close to the onset of Big Bang Nucleosynthesis (BBN). The initial abundance ratio is scanned over the interval $10^{-4} \leq r_\phi \leq 1$. Please note that solid contours correspond to positive values, while dashed contours correspond to negative values.

We observe that for the matter-like case, $w_\phi = 0$, as well as for the near-matter case, $w_\phi = 0.1$, the dilution factor can become extremely large, reaching values as high as $\Delta_\phi \sim 10^8$ for $\Gamma_\phi = 10^{-24} \GeV$ and $r_\phi = 1$. By contrast, for the radiation-like and kination cases, $w_\phi = 1/3$ and $w_\phi = 1$, respectively, we find $\Delta_\phi < 1$. In these regimes, instead of dilution, the presence of the $\phi$ component leads to an enhancement of the gravitino abundance. Consequently, the corresponding contours of $\log \Delta_\phi$ take negative values.

In the following section, we study in more detail the relation between the dilution factor $\Delta_\phi$ and the enhancement of the peak reheating temperature. Before proceeding, however, it is useful to develop a qualitative understanding of the dilution mechanism underlying the numerical results presented above.
Such an understanding can be obtained from the scaling behavior of the extra scalar  and its impact on the expansion history of the Universe.

The parameter $w_\phi$ determines the redshift behavior of the scalar energy density,
$\rho_\phi \propto a^{-3(1+w_\phi)}$,
to be compared with the radiation scaling $\rho_R \propto a^{-4}$. Consequently, for
$w_\phi < \frac{1}{3}$,
the scalar component redshifts more slowly than radiation and therefore its relative contribution to the total energy density increases during the cosmological evolution. In this case, $\phi$ may temporarily dominate the expansion of the Universe before decaying. The subsequent decay injects entropy into the thermal bath and dilutes the gravitino abundance previously generated through freeze-in processes. The smaller the value of $w_\phi$, the stronger this effect becomes, with the matter-like case, $w_\phi=0$, leading to the largest dilution factors.

On the other hand, for
$w_\phi > \frac{1}{3}$,
the scalar component redshifts faster than radiation and rapidly becomes subdominant. In this regime, entropy production is suppressed and the dominant effect arises from the modification of the Hubble expansion rate during gravitino production. This typically results in an enhancement of the gravitino abundance relative to the standard radiation-dominated scenario. The radiation-like case, $w_\phi=1/3$, approximately separates these two qualitatively distinct regimes.

The decay width $\Gamma_\phi$ controls the lifetime of the scalar component,
$\tau_\phi \sim \Gamma_\phi^{-1}$.
Smaller values of $\Gamma_\phi$ correspond to longer-lived scalar configurations, allowing $\phi$ to remain cosmologically relevant for an extended period. This increases the possibility of a transient $\phi$-dominated era and consequently enhances the entropy released during the decay process, leading to stronger dilution of the gravitino abundance. Conversely, sufficiently large values of $\Gamma_\phi$ imply an early decay of $\phi$, reducing its impact on the thermal history of the Universe.

Finally, the parameter $r_\phi$ determines the initial importance of the scalar component relative to radiation at reheating. Larger values of $r_\phi$ imply that $\phi$ becomes dynamically relevant earlier in the cosmological evolution, thereby amplifying both its effect on the expansion rate and the resulting entropy production. One therefore expects the dilution factor to increase with increasing $r_\phi$.

These simple scaling arguments qualitatively explain the numerical behavior observed in our analysis. Large dilution factors are obtained for small values of $w_\phi$, small decay widths $\Gamma_\phi$, and sufficiently large initial abundances $r_\phi$, whereas for $w_\phi \gtrsim 1/3$ the dilution effect disappears and may instead be replaced by an enhancement of the gravitino abundance.
Motivated by these considerations, in the following section we focus on the impact of the dilution factor, $\Delta_\phi$, on the peak reheating temperature, $\Trp$.

\section{The effect of  $\boldsymbol{\phi}$  on   $\boldsymbol{\Trp}$ }
\label{sect:num}
In this section, we investigate in detail the impact of the dilution factor $\Delta_\phi$ on the peak reheating temperature, $\Trp$. As discussed in the previous sections, the presence of the extra scalar  density can substantially modify the gravitino relic abundance through entropy production and alterations of the cosmological expansion rate. Consequently, the reheating temperature compatible with the observed gravitino DM abundance may differ significantly from the standard radiation-dominated prediction.

In this study, we impose the conservative upper bound
$T_{\rm reh}^{\rm peak} \lesssim 10^{16}~\GeV$. 
This scale is close to the characteristic energy scale of inflation suggested by typical GUT or high-scale inflationary scenarios. Reheating temperatures substantially above this value would approach, or even exceed, the energy density available at the end of inflation, thereby challenging the consistency of the effective cosmological description adopted in this work. Moreover, such extremely large reheating temperatures would generally require a more complete treatment of ultraviolet effects, including possible  Gravitational corrections and details of the reheating dynamics beyond the scope of our study.

As already anticipated in the previous section, the strongest dilution effects occur for small values of the equation-of-state parameter $w_\phi$, small decay widths $\Gamma_\phi$, and sufficiently large initial abundances $r_\phi$. Physically, these choices correspond to a long-lived scalar component whose energy density redshifts more slowly than radiation and therefore becomes increasingly important during the cosmological evolution. The subsequent decay of $\phi$ injects a large amount of entropy into the thermal bath, strongly diluting the previously generated gravitino abundance.

\begin{table*}[t!]
\begin{center}
\small
$\begin{array}{|c||c|c|c||c|c|c|}
\hline
&
\multicolumn{3}{c||}{w_\phi=0.1,\ \Gamma_\phi=10^{-16}\,{\rm GeV}}
&
\multicolumn{3}{c|}{w_\phi=1,\ \Gamma_\phi=10^{-12}\,{\rm GeV}}
\\
\cline{2-7}
M_{1/2}({\rm TeV})
&
m_{3/2}({\rm GeV})
&
T_{\rm reh}^{\rm peak,\phi}({\rm GeV})
&
\Delta_\phi
&
m_{3/2}\,({\rm GeV})
&
T_{\rm reh}^{\rm peak,\phi}({\rm GeV})
&
\Delta_\phi
\\
\hline
\hline
1&645&1.45\times10^{15}&8.80\times10^{5}&763&3.64\times10^{8}&3.40 \times10^{-1}
\\
\hline
2&1441&1.84 \times10^{14}&2.37\times10^{5}&1584&1.77\times10^{8}&3.40\times10^{-1}
\\
\hline
5&3826&1.22 \times10^{13}&4.20 \times10^{4}&3896&6.77\times10^{7}&3.39\times10^{-1}
\\
\hline
10&7748&1.58\times10^{12}&1.14\times10^{4}&7931&3.27\times10^{7}&3.38\times10^{-1}
\\
\hline
\end{array}$
\caption{\it
The values of $m_{3/2}$, the peak reheating temperature
$T_{\rm reh}^{\rm peak,\phi}$, and the dilution factor $\Delta_\phi$
for various choices of $M_{1/2}$, assuming complete gravitino DM,
$\Omega_{3/2}h^2=0.12$, in the modified cosmological background.
We compare two representative cases:
a near matter-like component with $w_\phi=0.1$ and
$\Gamma_\phi=10^{-16}\GeV$, and a kination-dominated component with
$w_\phi=1$ and $\Gamma_\phi=10^{-12}\GeV$, both for
$r_\phi=10^{-3}$.
The dilution factor is defined as
$\Delta_\phi \equiv
\Omega_{3/2} /\Omega_{3/2}^{\phi}$
evaluated at the position of the shifted reheating peak.}
\label{table:Tehhmax_phi_combined}
\end{center}
\end{table*}

Since the gravitino relic density scales approximately linearly with the reheating temperature, large dilution factors can be compensated by correspondingly larger values of $T_{\rm reh}$. As a result, in regions of parameter space where the dilution becomes extremely efficient, one obtains reheating temperatures satisfying $T_{\rm reh}^{\rm peak} \gtrsim 10^{16}\GeV$. 
Such parameter regions are excluded from the phenomenological analysis presented below. In practice, this mainly affects configurations close to the matter-dominated limit, characterized by very small values of $w_\phi$, together with small decay widths $\Gamma_\phi$ and large initial abundances $r_\phi$.

A useful analytic estimate of the dilution effect can be obtained by considering the evolution of the additional scalar  before it decays. If gravitino production is completed before the decay of $\phi$, the subsequent evolution reduces the final gravitino yield. In this case, the dilution factor defined in Eq.~(\ref{Eq:delta}),
\begin{equation}
\Delta_\phi
=\frac{Y_{3/2}}{Y_{3/2}^{\phi}},
\end{equation}
provides a direct measure of the suppression of the gravitino abundance.
Assuming that the decay occurs when $H\simeq \Gamma_\phi$, the corresponding decay temperature is approximately
\begin{equation}
T_\phi
\simeq
\left(
\frac{90}{\pi^2 g_*(T_\phi)}
\right)^{1/4}
\sqrt{\Gamma_\phi M_P}\, .
\end{equation}
For a matter-like scalar component, $w_\phi=0$, the ratio of scalar to radiation energy densities evolves as
\begin{equation}
\frac{\rho_\phi}{\rho_R}
\simeq
r_\phi
\frac{T_{\rm reh}}{T},
\end{equation}
prior to the decay of $\phi$. Consequently, the scalar component starts to dominate the energy density at the temperature 
\begin{equation}
T_{\rm dom}
\simeq
r_\phi\,T_{\rm reh}.
\end{equation}
If a period of $\phi$ domination is realized, one expects the dilution factor to scale parametrically as
\begin{equation}
\Delta_\phi
\sim
\frac{T_{\rm dom}}{T_\phi}
\sim
\frac{r_\phi T_{\rm reh}}{T_\phi}.
\label{Eq:dil_app}
\end{equation}

Equation~(\ref{Eq:dil_app}) explains the main trends observed in our numerical analysis. The dilution increases with the initial abundance $r_\phi$ and decreases with the decay temperature $T_\phi$, or equivalently with increasing decay width $\Gamma_\phi$, since $T_\phi \propto \sqrt{\Gamma_\phi}$. As a result, larger values of $r_\phi$ and smaller values of $\Gamma_\phi$ lead to a stronger suppression of the gravitino abundance. For a fixed relic density, this suppression can be compensated by increasing the reheating temperature, suggesting the approximate scaling
\begin{equation}
T_{\rm reh}^{\rm peak,\phi}
\sim
\Delta_\phi\,T_{\rm reh}^{\rm peak},
\label{Eq:dil_estim}
\end{equation}
up to corrections associated with the modified expansion rate during gravitino production.
In the following, Eqs.~(\ref{Eq:dil_app}) and (\ref{Eq:dil_estim}) will serve as useful guides for interpreting the numerical results and understanding the dependence of the shifted peak reheating temperature $T_{\rm reh}^{\rm peak,\phi}$ on the parameters of the $\phi$-sector.

\begin{figure}[t!]
\centering

\includegraphics[width=0.48\textwidth]{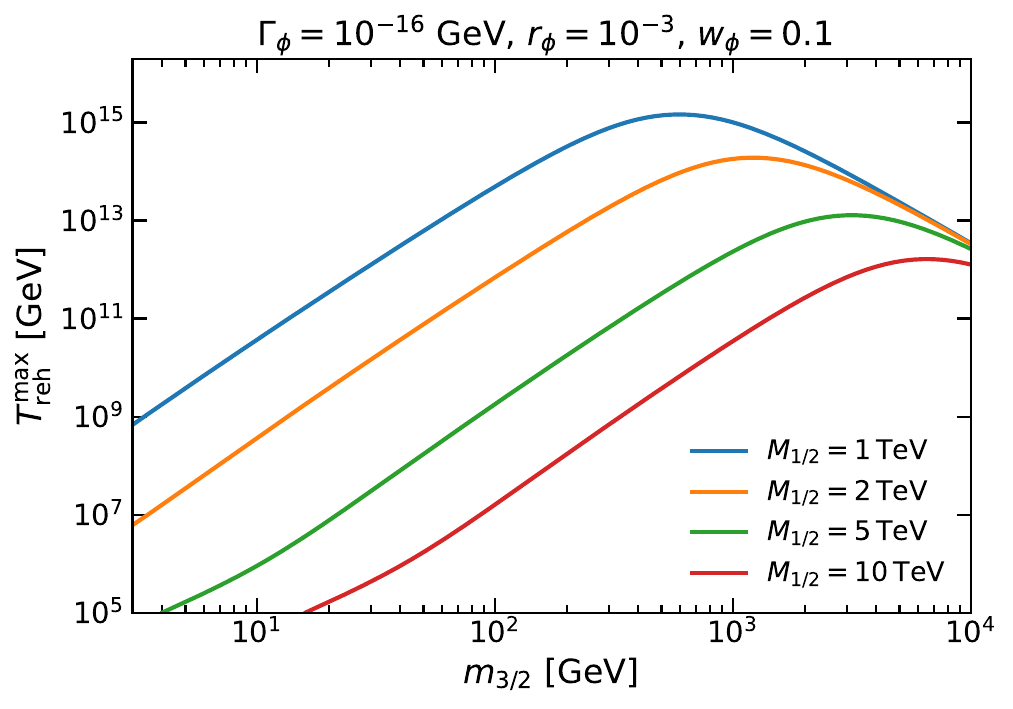}  
\includegraphics[width=0.48\textwidth]{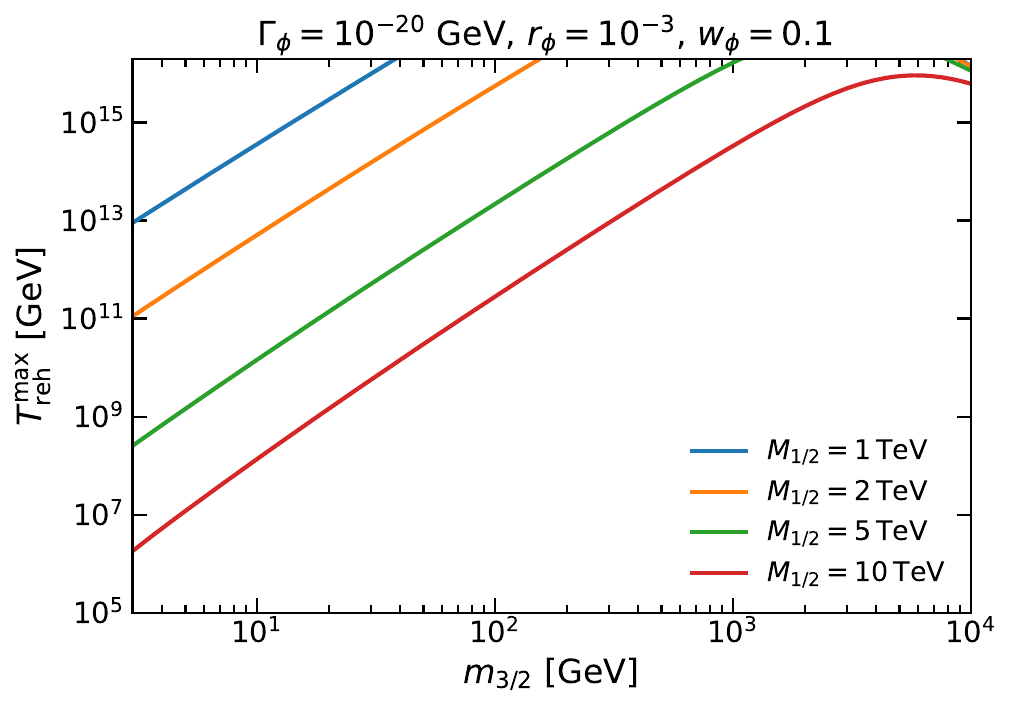} \\[0.5cm]
\includegraphics[width=0.48\textwidth]{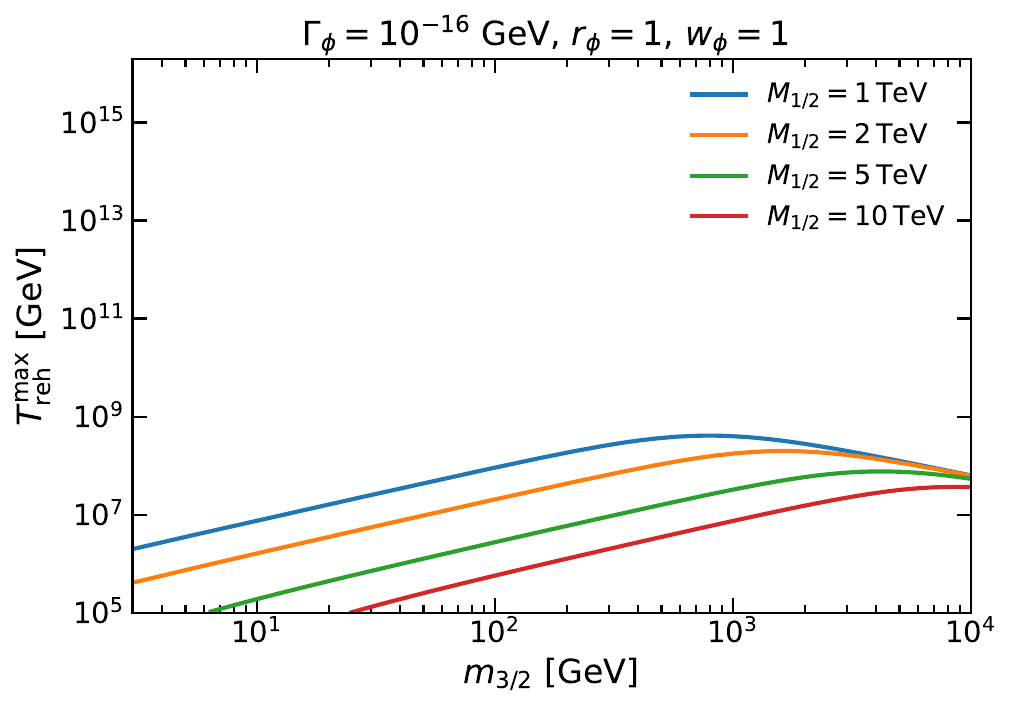}
\includegraphics[width=0.48\textwidth]{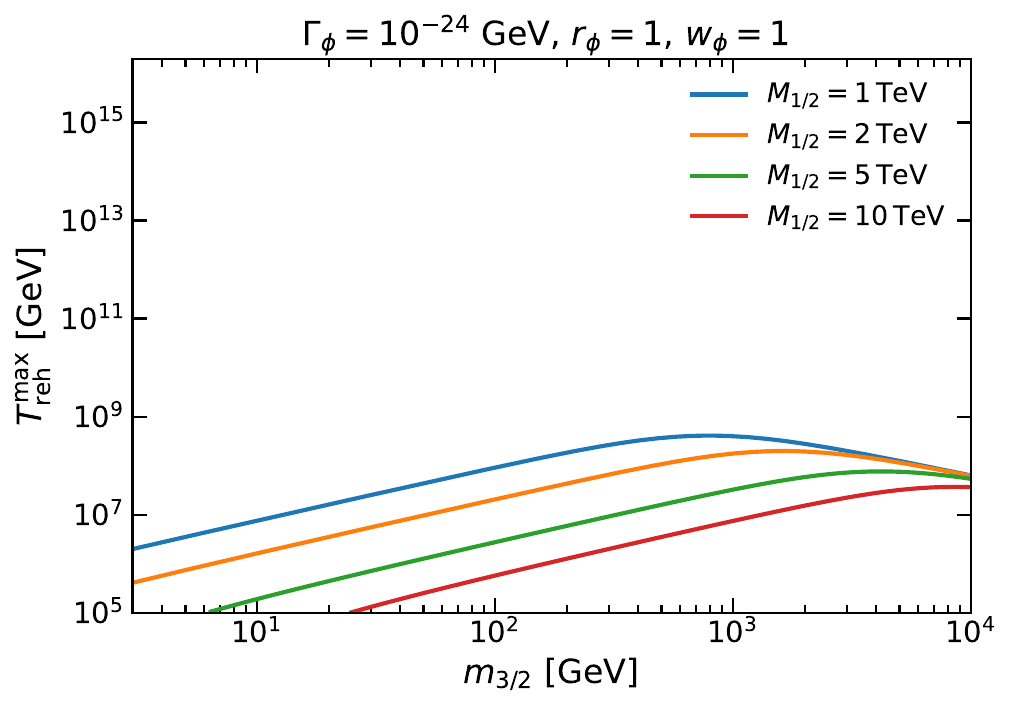}
\caption{\it  The curves of fixed $\Omega_{3/2}h^2=0.12$ in the $m_{3/2},T_{\rm reh}^{\rm max}$ 
plane using the exact numerical 
solution of the gravitino abundance Boltzmann equation in the non-standard background,
using the values of  $r_\phi $, $w_\phi $ and $\Gamma_\phi $ as displayed in the figures. }
\label{fig:dilut_phi}
\end{figure}

In Table~\ref{table:Tehhmax_phi_combined}, we summarize the values of the gravitino mass $m_{3/2}$, the shifted peak reheating temperature $T_{\rm reh}^{\rm peak,\phi}$, and the corresponding dilution factor $\Delta_\phi$ for several benchmark choices of $M_{1/2}$, under the assumption that gravitinos account for the total DM abundance, $\Omega_{3/2}h^2=0.12$.
Two representative cosmological scenarios are compared. The first corresponds to a near matter-dominated scalar component with $w_\phi=0.1$ and $\Gamma_\phi=10^{-16}\GeV$, while the second corresponds to a kination-dominated configuration with $w_\phi=1$ and $\Gamma_\phi=10^{-12}\GeV$. In both cases, we fix the initial abundance ratio to $r_\phi=10^{-3}$.
The table clearly illustrates the qualitatively different impact of the two regimes. In the near matter-like case, substantial entropy production leads to very large dilution factors and consequently to a significant enhancement of the peak reheating temperature. By contrast, in the kination case, the gravitino abundance is enhanced rather than diluted, resulting in $\Delta_\phi<1$ and correspondingly smaller values of $T_{\rm reh}^{\rm peak,\phi}$. Another interesting feature for $w_\phi=1$ is that the dilution factor remains nearly independent of $\moh$.

An analogous behavior is illustrated in Fig.~\ref{fig:dilut_phi}, where we present four panels showing the contours corresponding to fixed $\Omega_{3/2} h^2=0.12$ in the $(m_{3/2},T_{\rm reh})$ plane. The contours are obtained from the exact numerical solution of the gravitino Boltzmann equation in the modified cosmological background.
In the upper panels, we consider the near matter-like case with $r_\phi=10^{-3}$ and $w_\phi=0.1$. The left and right panels correspond to $\Gamma_\phi=10^{-16}\GeV$ and $\Gamma_\phi=10^{-20}\GeV$, respectively. In both cases, contours are shown for the benchmark values $\moh=1$, $2$, $5$, and $10~\TeV$. We observe that for $\Gamma_\phi=10^{-16}\GeV$ the shifted reheating peak for $\moh=1~\TeV$ already approaches the typical inflationary scale, while for $\Gamma_\phi=10^{-20}\GeV$ it exceeds the upper bound adopted in our analysis, namely $T_{\rm reh}^{\rm peak,\phi}\lesssim10^{16}\GeV$.
In addition, comparing with Fig.~\ref{fig:grav_ns}, corresponding to $\Gamma_\phi=10^{-12}\GeV$, we observe that decreasing $\Gamma_\phi$ by four orders of magnitude leads to an approximately proportional increase of $\Trpf$ by the same factor, as anticipated from the analytic considerations discussed earlier.

The lower panels correspond to a kination-dominated configuration with $r_\phi=1$ and $w_\phi=1$. The left and right panels show the cases $\Gamma_\phi=10^{-16}\GeV$ and $\Gamma_\phi=10^{-24}\GeV$, respectively. In contrast to the upper panels, the presence of the $\phi$ component now enhances rather than dilutes the gravitino abundance, corresponding to $\Delta_\phi<1$. Consequently, the reheating peaks are shifted toward lower values relative to the standard radiation-dominated case, even for decay widths close to the onset of BBN.

\begin{figure}[t!]
\centering
\includegraphics[width=0.49\textwidth]{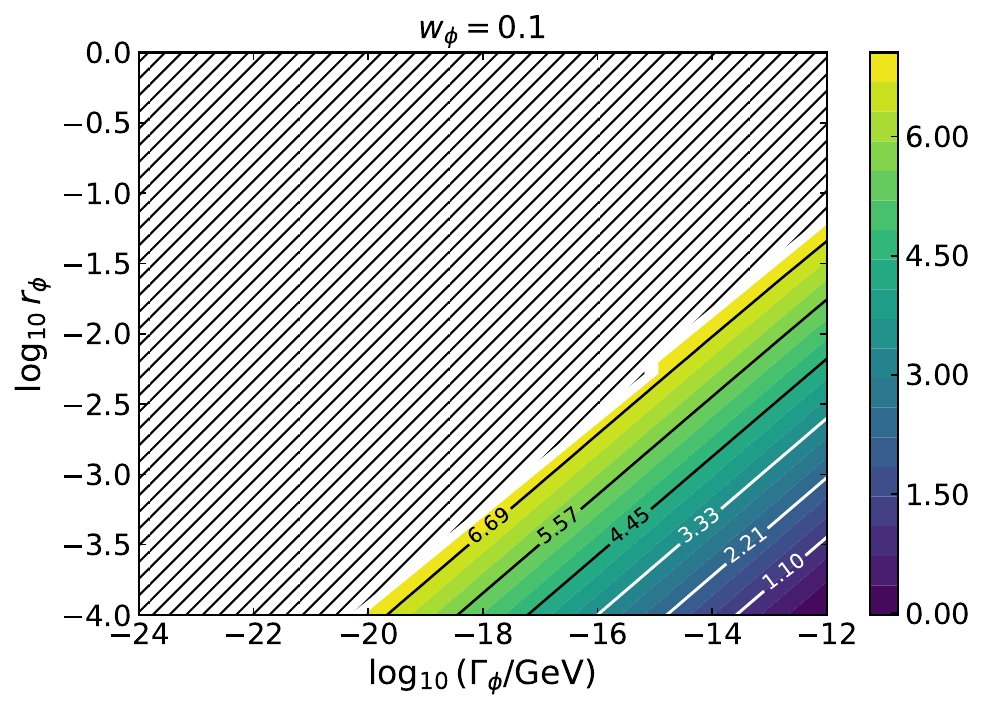}
\includegraphics[width=0.49\textwidth]{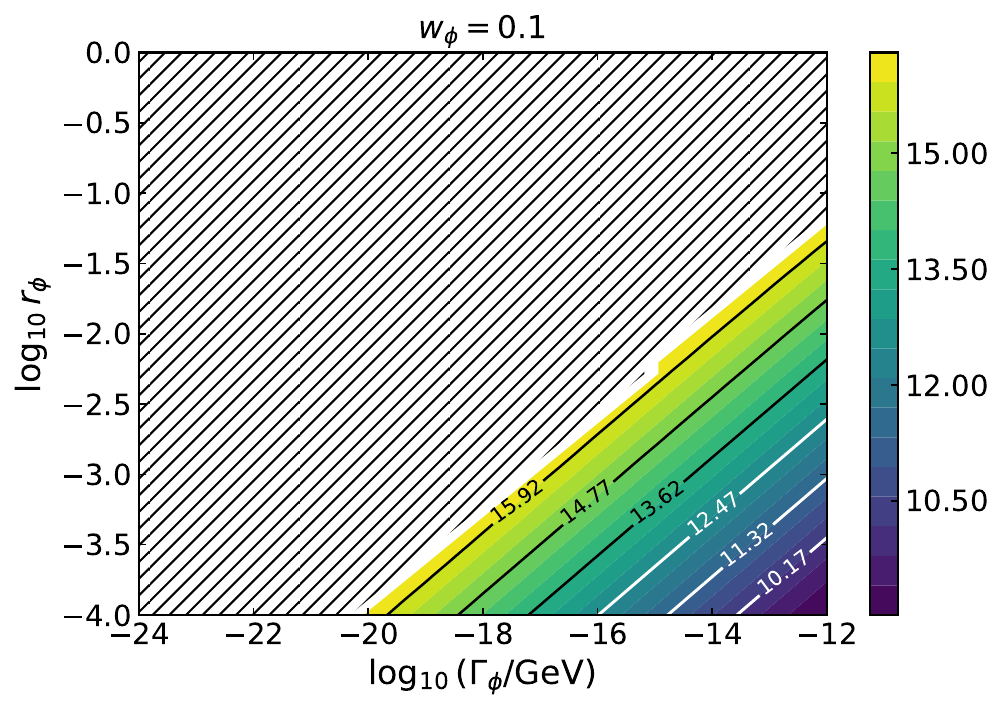}\\[0.5cm]
\includegraphics[width=0.49\textwidth]{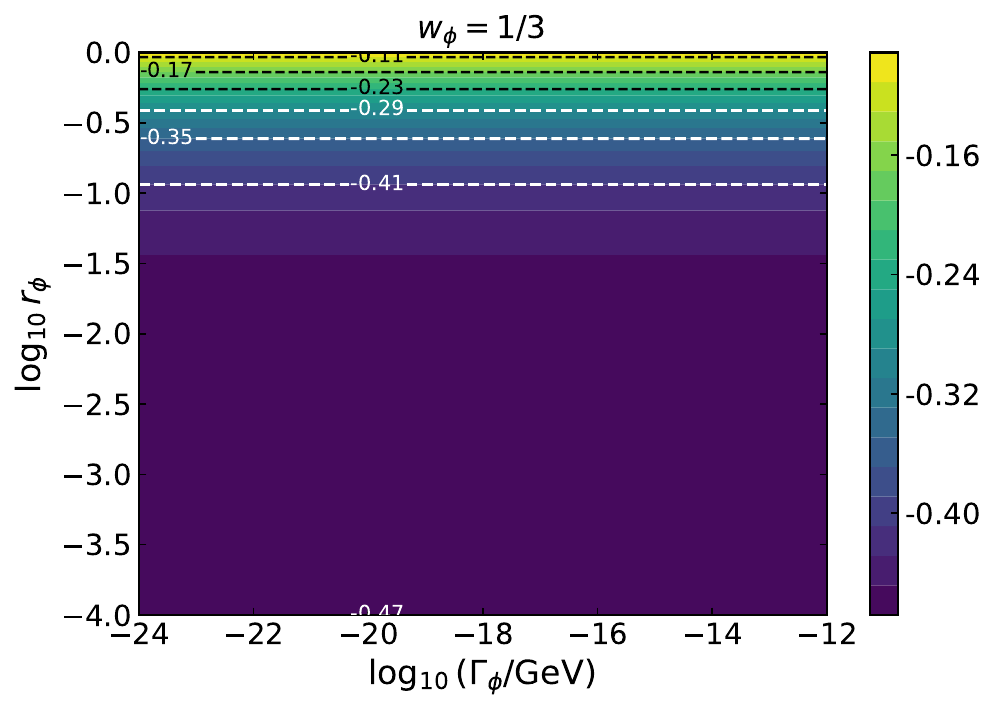}
\includegraphics[width=0.49\textwidth]{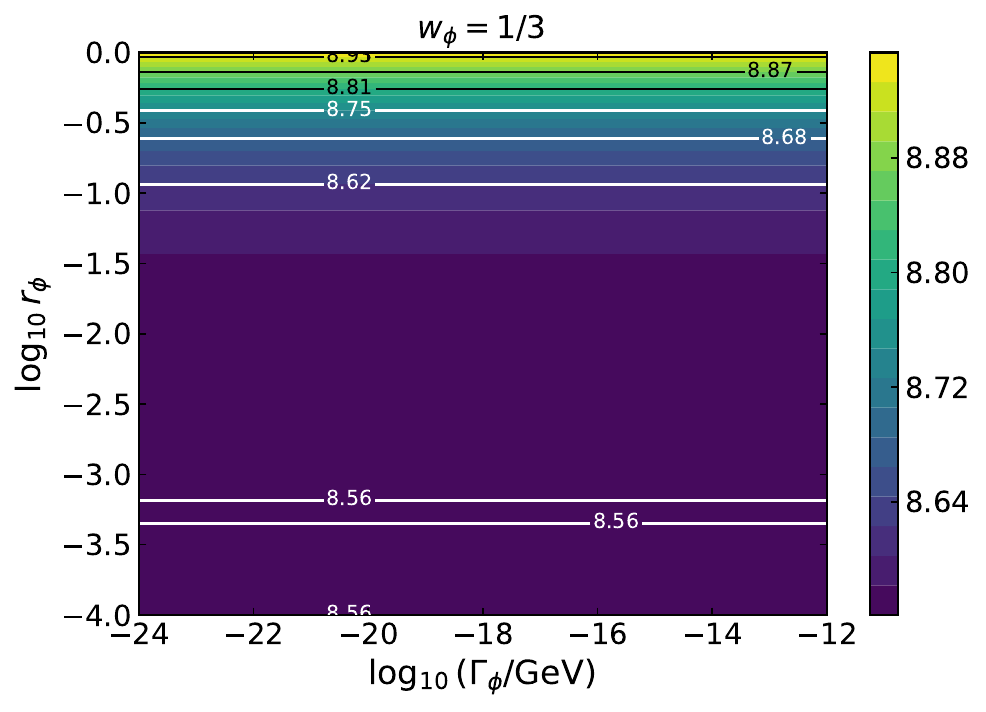}\\[0.5cm]
\includegraphics[width=0.49\textwidth]{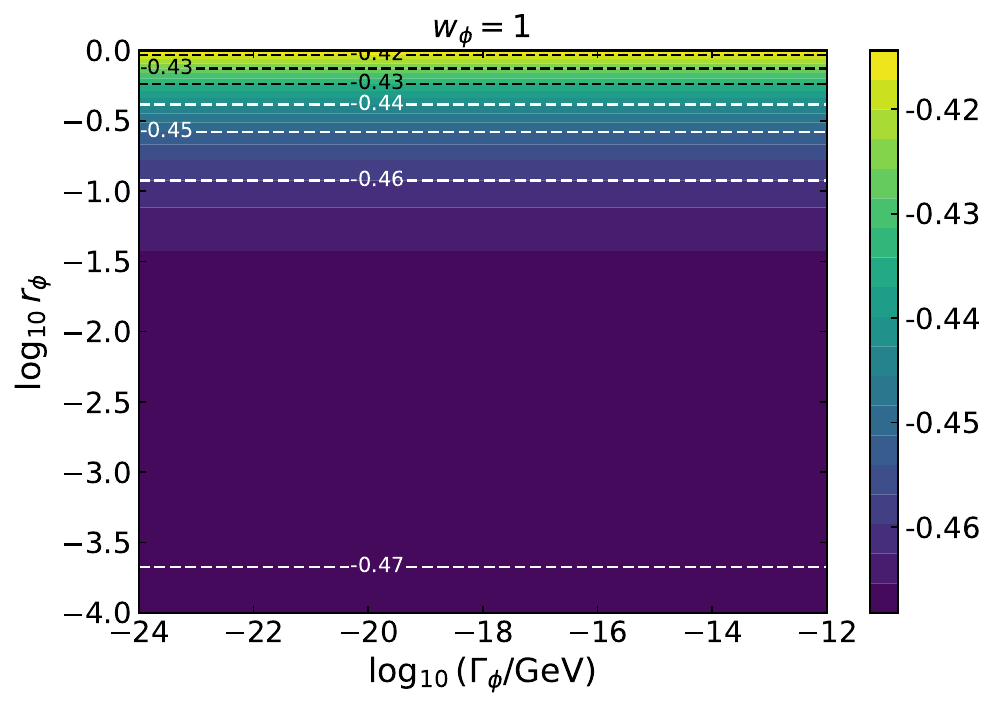} 
\includegraphics[width=0.49\textwidth]{ 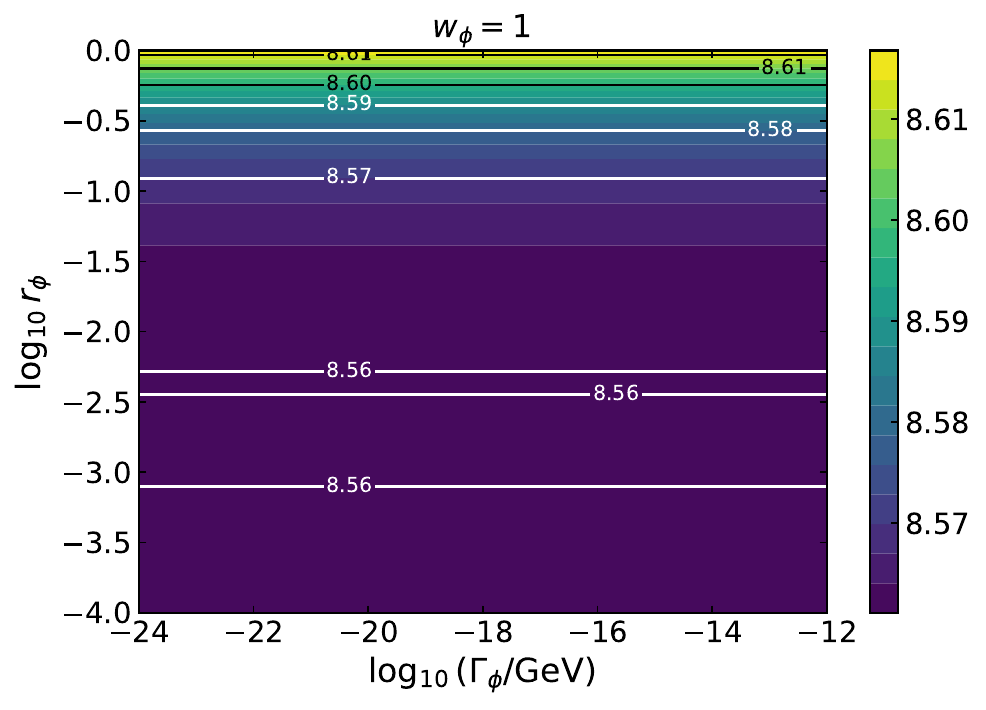} 
\caption{\it  The dilution factor at the shifted reheating peak, $\Delta_\phi$ (left panels), and the corresponding peak reheating temperature, $T_{\rm reh}^{\rm peak,\phi}$ (right panels), shown in the $(\Gamma_\phi,r_\phi)$ plane for $M_{1/2}=1~\TeV$. The three rows correspond to $w_\phi=0.1$, $w_\phi=1/3$, and $w_\phi=1$, respectively. In the shaded regions we get   $T_{\rm reh}^{\rm peak,\phi} >  10^{16} \GeV$. }
\label{fig:Treh_gamma}
\end{figure}

For the numerical determination of the shifted reheating peak, $T_{\rm reh}^{\rm peak,\phi}$, we employ an efficient strategy based on the corresponding peak obtained in the standard radiation-dominated cosmology. The location of the standard peak provides an excellent initial estimate for the nonstandard case and significantly reduces the computational cost of the scan.

Starting from the standard peak configuration, we first evaluate whether the presence of the $\phi$ component suppresses or enhances the gravitino abundance. In the dilution regime, where the gravitino abundance is reduced, the corresponding nonstandard contour is shifted toward larger reheating temperatures. Conversely, in the enhancement regime, the contour moves toward smaller reheating temperatures.

Using this information, we numerically reconstruct the contour corresponding to the observed gravitino DM abundance, $\Omega_{3/2}^{\phi} h^2 = 0.12$, in the $(m_{3/2},T_{\rm reh})$ plane. The contour is then followed iteratively until the maximum reheating temperature is identified. At this point, we evaluate the corresponding dilution factor $\Delta_\phi$, which quantifies the modification of the gravitino abundance relative to the standard cosmological scenario.

Using the analysis described above, Fig.~\ref{fig:Treh_gamma} presents a six-panel summary of the impact of the additional $\phi$ component on the dilution factor $\Delta_\phi$ and on the shifted reheating-temperature peak $T_{\rm reh}^{\rm peak,\phi}$ for $\moh=1~\TeV$. The dilution factor is evaluated at the position of the shifted peak in each case.
The left column shows contours of $\log \Delta_\phi$ in the $(\Gamma_\phi,r_\phi)$ plane, while the right column displays the corresponding contours of $\log T_{\rm reh}^{\rm peak,\phi}$. The three rows correspond to the representative choices $w_\phi=0.1$, $w_\phi=1/3$, and $w_\phi=1$, respectively. As in Fig.~\ref{fig:dil_1tev}, solid contours correspond to positive values, while dashed contours correspond to negative.

Parameter regions violating this condition are indicated by the shaded areas. For the exact matter-dominated case, $w_\phi=0$, the dilution effect becomes so efficient that essentially the entire parameter space explored would lead to reheating temperatures above this bound. For this reason, the corresponding plots are omitted.

A partially excluded region is also visible in the upper-left corner of the $w_\phi=0.1$ panels. In this regime, small decay widths together with sufficiently large initial abundances generate extremely large entropy production, pushing the reheating peak beyond the imposed upper limit. More specifically, the approximate boundary of the allowed region is determined by configurations around $r_\phi \sim 10^{-4}$ and $\Gamma_\phi \sim 10^{-20}\GeV$, as well as $r_\phi \sim 10^{-1}$ and $\Gamma_\phi \sim 10^{-12}\GeV$. Along this boundary, the dilution factor reaches values of order $\Delta_\phi \sim 10^7$, in qualitative agreement with the analytic estimate of Eq.~(\ref{Eq:dil_estim}).
By contrast, for the radiation-like and kination cases, $w_\phi=1/3$ and $w_\phi=1$, respectively, the dilution factor becomes smaller than unity throughout most of the parameter space. Consequently, the gravitino abundance is enhanced rather than diluted, leading to reduced values of $T_{\rm reh}^{\rm peak,\phi}$ compared to the standard radiation-dominated scenario.

The results presented in this section show that the presence of the extra  scalar  can significantly modify the reheating-temperature bounds associated with gravitino freeze-in DM. Depending on the values of $(w_\phi,\Gamma_\phi,r_\phi)$, the decay of $\phi$ may either dilute or enhance the gravitino abundance relative to the standard radiation-dominated case.
The benchmark results summarized in Table~\ref{table:Tehhmax_phi_combined}, together with the contour plots in Figs.~\ref{fig:dilut_phi} and \ref{fig:Treh_gamma}, illustrate this behavior. For small values of $w_\phi$, corresponding to near matter domination, substantial entropy production can generate very large dilution factors and shift the reheating peak toward significantly larger values.
By contrast, for radiation-like or kination-like configurations, the gravitino abundance is enhanced rather than diluted, leading to lower values of the reheating peak.
Thus,  even a subdominant additional cosmological component during the freeze-in era can substantially affect the gravitino relic abundance and the resulting constraints on the reheating temperature.

\section{Summary and Perspectives}
\label{sect:recap}
In this work, we extended the standard radiation-dominated treatment of thermal gravitino production by introducing an additional energy component, $\rho_\phi$. Such scalar fields arise naturally in supergravity and string-inspired frameworks. Examples include moduli associated with compactification sectors, flat directions of supersymmetric theories, saxion fields related to the axion supermultiplet, as well as hidden-sector scalars interacting only gravitationally, or very weakly, with the visible sector.
In a broader framework, $\phi$ may represent any long-lived scalar condensate whose energy density becomes cosmologically relevant after reheating and before the onset of the standard radiation-dominated era. Depending on its equation of state and decay properties, this component can modify the expansion history of the Universe, generate entropy production, and consequently alter the freeze-in production of gravitinos.

Our analysis remained agnostic regarding the microscopic origin and dynamical nature of $\phi$. We therefore modeled it phenomenologically as a decaying fluid characterized by an equation-of-state parameter $w_\phi$, a decay width $\Gamma_\phi$, and an initial energy density $\rho_\phi(T_{\rm reh})$. The cosmological framework considered in this work thus consists of three interacting components: radiation, gravitinos, and the scalar component $\phi$.

We systematically explored the extended parameter space associated with $\phi$, using as benchmark points the maxima of the reheating temperature, $T_{\rm reh}^{\rm peak}$, obtained in the conventional radiation-dominated scenario. Our analysis shows that the effect of the extra field  depends crucially on the value of the barotropic parameter $w_\phi$. In particular, for
$0 \leq w_\phi < \frac{1}{3}$,
the presence of $\phi$ leads to a dilution of the gravitino abundance, whereas for
$\frac{1}{3} \leq w_\phi \leq 1$,
the gravitino abundance is enhanced relative to the standard case.

We scanned values of the decay width in the range
$10^{-24} \GeV \lesssim \Gamma_\phi \lesssim 10^{-12} \GeV$, 
corresponding to lifetimes extending up to $\mathcal{O}(10^{-1}){\rm s}$, close to the onset of BBN. We found that sufficiently small values of $\Gamma_\phi$, particularly $\Gamma_\phi \lesssim 10^{-16}\GeV$, can produce extremely large dilution factors, resulting in shifted peak reheating temperatures satisfying
$T_{\rm reh}^{\rm peak,\phi} \gtrsim 10^{16} \GeV$.
Since such values approach the scale of inflation and the limits of validity of our effective treatment, we did not pursue these regions further in the present analysis.

For the initial abundance of the scalar component, we considered values of
$10^{-4} \lesssim r_\phi \equiv \frac{\rho_\phi}{\rho_R} \lesssim 1$.
We found that in the matter-like case, $w_\phi = 0$, essentially the entire parameter space explored leads to very large dilution effects and consequently to
$T_{\rm reh}^{\rm peak,\phi} \gtrsim 10^{16}\GeV$.
For moderately larger values, such as $w_\phi = 0.1$, the dilution effect becomes less dramatic, and values satisfying
$T_{\rm reh}^{\rm peak,\phi} \lesssim 10^{16} \GeV$
can still be obtained for sufficiently large decay widths and/or small initial abundances of $\phi$.

By contrast, for the radiation-like case, $w_\phi = 1/3$, and even more strongly for the kination case, $w_\phi = 1$, the additional component enhances the gravitino relic abundance instead of diluting it. As a consequence, the corresponding reheating temperature satisfies
$T_{\rm reh}^{\rm peak,\phi} < T_{\rm reh}^{\rm peak}$,
thereby strengthening the cosmological bounds on the reheating scale. The same qualitative behavior persists throughout the interval
$\frac{1}{3} < w_\phi \leq 1$.

Freeze-in models provide a viable and attractive alternative to conventional DM scenarios, naturally evading the stringent direct and indirect experimental constraints that apply to more strongly interacting DM candidates. Our results demonstrate that the presence of an additional cosmological component during the freeze-in production era can substantially modify the cosmological predictions associated with feebly interacting DM particles.

A final remark concerns the implications of the additional scalar component for baryogenesis. If the baryon asymmetry is generated before the decay of $\phi$, the associated entropy injection is expected to affect the final baryon asymmetry, and consequently the reheating-temperature requirements of scenarios such as thermal leptogenesis. However, the impact of the $\phi$ component is generally more subtle than a simple dilution effect. In particular, the modified expansion history, the timing of the $\phi$ decay, and the interplay between particle production and entropy injection can all influence the final abundances. Indeed, in~\cite{Dalianis:2023ixz} demonstrated,  within a simple freeze-in cogenesis framework based on dimension-five operators analogous to those relevant for the gravitino production, that an additional scalar component can significantly modify the predicted dark matter and baryon abundances compared to the standard cosmological scenario.

A quantitative assessment of baryogenesis in the framework considered here would require a simultaneous treatment of the evolution of the lepton (or baryon) asymmetry together with the scalar sector for a specific baryogenesis mechanism, and lies beyond the scope of the present work. Nevertheless, our results demonstrate that the gravitino upper bound on the reheating temperature can be relaxed substantially. As a consequence, gravitino DM scenarios can accommodate significantly higher reheating temperatures, alleviating the tension between thermal gravitino production and high-scale baryogenesis mechanisms such as thermal leptogenesis, even in the presence of future collider constraints favoring heavier gaugino masses. A more detailed investigation of specific microscopic realizations of the scalar sector, together with a comprehensive study of their implications for baryogenesis and other cosmological observables, is left for future work.

\end{document}